\documentclass[aps,pre,reprint,superscriptaddress]{revtex4-2}
\usepackage{graphicx}
\graphicspath{{figures/}}
\usepackage[T1]{fontenc} 
\usepackage{lmodern}  
\usepackage{hyperref}
\usepackage{csquotes}
\usepackage{amsmath}
\usepackage{amsfonts} 
\usepackage{amssymb}
\usepackage{textcmds}
\usepackage{dsfont}

\usepackage[usenames,dvipsnames]{xcolor}
\hypersetup{colorlinks=true, linkcolor=BrickRed, urlcolor=blue!50!black, citecolor=blue!50!black}

\usepackage{amsmath,amsthm,amsfonts,amssymb,times,bbm,graphicx,color,epsfig,bm}
\usepackage{url} 
\usepackage{hyperref}
\renewcommand{\vec}[1]{\textbf{\textrm{#1}}}

\usepackage[normalem]{ulem}

\DeclareMathAlphabet\mathbfcal{OMS}{cmsy}{b}{n}

\begin{document}

\noindent\small Phys. Rev. E 113, 035305 (2026). \\
\vfill 

\title{When velocity autocorrelations mirror force autocorrelations: Exact noise-cancellation in interacting Brownian systems}
\author{Anton L\"uders}
\email{anton.lueders@uibk.ac.at}
\affiliation{Institut f\"ur Theoretische Physik, Universit\"at Innsbruck, Technikerstra{\ss}e, 21A, A-6020 Innsbruck, Austria}
\author{Suvendu Mandal}
\affiliation{Institut f\"ur Physik kondensierter Materie, Technische Universit\"at Darmstadt, Hochschulstra{\ss}e 8, D-64289 Darmstadt, Germany}
\author{Thomas Franosch}
\email{thomas.franosch@uibk.ac.at}
\affiliation{Institut f\"ur Theoretische Physik, Universit\"at Innsbruck, Technikerstra{\ss}e, 21A, A-6020 Innsbruck, Austria}

\begin{abstract}

Resolving the mean-squared displacement (MSD) and velocity autocorrelation function (VACF) of interacting Brownian particles remains a central challenge in simulations of soft-matter systems, especially at low densities where particle-particle interactions are sparse and signals are dominated by thermal noise. A recently proposed noise-cancellation (NC) algorithm [Mandal et al.\@ Phys.\@ Rev.\@ Lett.\@ 123, 168001 (2019)] addresses this by decomposing particle trajectories into two components: free Brownian motion and interaction-induced displacements. The NC approximation enhances signal clarity by neglecting cross-correlations between the total particle displacements and the extracted interaction-induced contributions of the trajectories --- an assumption that has so far lacked rigorous theoretical justification. In this work, we establish an exact theoretical relation between the VACF, the force autocorrelation function (FACF) characterizing the interaction-induced contributions, and these cross-correlations, which is valid for Brownian systems. We show that in thermal equilibrium, the cross-correlations vanish for Brownian systems because the VACF is strictly proportional to the negative FACF, which establishes the NC algorithm as an exact method. In contrast, for Brownian nonequilibrium systems, the cross-correlations remain finite, providing a direct fingerprint of nonequilibrium physics in such systems and a criterion to distinguish equilibrium from nonequilibrium states. Here, suitable corrections must be applied for the NC method to remain accurate. Our results expand the scope of the NC algorithm to a broad range of soft-matter systems in and out of equilibrium, where it has the potential to strongly enhance the resolution of VACF data obtained through simulations in future studies.

\end{abstract}

\maketitle

\section{Introduction}

Understanding how microscopic fluctuations and their correlations decay over time is essential to characterize the transport properties of liquids, complex fluids, and colloidal suspensions. A central quantity in this context is the velocity autocorrelation function (VACF), which measures how the velocity of a tagged particle remains correlated with itself over time. The VACF not only reveals the underlying dynamical processes at play but also connects directly to macroscopic transport coefficients, such as the diffusivity, via the Green-Kubo relations~\cite{Mouas2012, Heptner2015, Hansen2013, Kubo2012}.

In simple molecular fluids, the VACF encodes rich dynamical information across time scales. Classical kinetic theory predicts an exponential decay of the VACF in dilute gases due to uncorrelated binary collisions~\cite{Graham1972, Cichocki1994}. However, groundbreaking simulations by Alder and Wainwright revealed a drastic departure from this behavior: in Newtonian many-body systems, the VACF exhibits a long-time algebraic decay as $t^{-3/2}$ in three dimensions~\cite{Alder1970}. This long-time tail, now firmly established both theoretically and experimentally~\cite{Ernst1970, Dorfman1970, Lukic2005, Jeney2008, Franosch2011, Huang2011}, arises from the hydrodynamic backflow resulting from momentum conservation --- a hallmark of collective fluid behavior.

In colloidal systems, by contrast, this picture changes fundamentally. Due to strong viscous damping from the surrounding solvent, momentum is rapidly dissipated and no longer conserved at the particle scale. Instead, long-time dynamics are governed by the particle-number conservation~\cite{Mandal2019}. This leads to a steeper but negative VACF decay, typically scaling as $t^{-5/2}$ in three dimensions~\cite{Hanna1981, Ackerson1982, Felderhof1983} [or $ t^{-(d+2)/2} $ in the $ d $-dimensional case]. Accurately resolving this long-time behavior is crucial for determining transport coefficients such as the diffusivity, yet extracting the VACF from stochastic simulations remains a formidable challenge.

The difficulty arises because the VACF signal decays rapidly and becomes increasingly buried in noise at long times, especially in overdamped dilute systems. As a result, achieving statistically converged estimates typically requires a prohibitively large number of independent simulations~\cite{Frenkel1987, Mandal2019, Ramirez2010}. This is a long-standing problem in computational statistical mechanics, where various strategies, such as fluctuation-subtraction techniques~\cite{Oettinger1994, Melchior1996, Wagner1997}, have been developed to reduce statistical variance. In the specific context of time-dependent correlation functions, several optimized sampling approaches have been proposed to enhance statistical efficiency~\cite{Siems2018, Ramirez2010, Dubbeldam2009, Frenkel1987}.

A promising recent advance is the noise-cancellation (NC) algorithm introduced by two of the current authors and collaborators~\cite{Mandal2019} for the calculation of the mean-squared displacement (MSD) and the VACF in Brownian dynamics (BD) simulations. Building on earlier ideas by Frenkel~\cite{Frenkel1987}, the NC method decomposes particle displacements into two components: the free Brownian motion and a contribution due to interparticle forces. By neglecting the cross-correlations between the total particle displacements and the extracted interparticle-force contributions, the method reconstructs the VACF from the second time derivative of the mean-squared displacement (MSD), drastically improving the signal-to-noise ratio at long times. This gain in signal is achieved because the remaining non-trivial terms of the MSD differ from zero only for direct particle-particle interactions, eliminating the need to average out the Brownian fluctuations~\cite{Rusch2024}. Accordingly, no explicit dependence on the Brownian noise persists. This strategy proved to increase the resolution of the long-time tails of the VACF by several orders of magnitude~\cite{Mandal2019}, and it was shown that numerical MSD results obtained by the NC procedure achieve a higher precision than standard algorithms when studying dilute or weakly interacting systems~\cite{Rusch2024}.

However, the theoretical basis for the NC approximation remains incomplete. While numerical evidence suggests that the ignored cross-correlation terms are indeed negligible in equilibrium and dilute conditions, the underlying reasons for this remain unclear. Moreover, in nonequilibrium or strongly interacting Brownian systems, these cross-correlations could be significant, potentially undermining the accuracy of the NC method.

In this work, we develop a rigorous theoretical foundation for the NC algorithm that is applicable to systems that can be described by overdamped Langevin equations (i.\@e.\@, Brownian systems). We derive an exact relation connecting the VACF, the force autocorrelation function (FACF), and the cross-correlation terms omitted in the NC method. Our key result is that, in thermal equilibrium, the NC algorithm is an exact expression for Brownian systems, meaning that the cross-correlation term always vanishes under such conditions. This result not only explains the remarkable accuracy of the NC method in BD simulations but also provides the first rigorous justification for its application to equilibrium soft-matter setups. In striking contrast, once Brownian systems are driven out of equilibrium, the cross-correlations remain finite and cannot be ignored. These non-vanishing contributions encode a distinct fingerprint of nonequilibrium physics, thereby offering a powerful tool for distinguishing equilibrium from nonequilibrium states. Lastly, we find that, even in nonequilibrium, the NC algorithm can sometimes be adjusted to give reasonable or even exact results, which expands the computation scheme of Ref.~\cite{Mandal2019} to Brownian nonequilibrium systems for the first time.

This work is structured as follows: First, we summarize the fundamentals of the NC algorithm as discussed in Refs.~\cite{Mandal2019, Rusch2024} in Sec.~\ref{sec:NCA}. Then, we derive the connection between the VACF, the FACF, and the cross-correlations in Brownian systems in Sec.~\ref{sec:Gen}. Afterwards, we conclude that the NC algorithm is exact in thermal equilibrium when applied to determine the VACF in Sec.~\ref{sec:EQ} and study the cross-correlations for Brownian nonequilibrium systems in Sec.~\ref{sec:NEQ}. In the sections addressing the Brownian equilibrium and nonequilibrium systems, respectively, we analyze multiple example systems to test our results and show that the NC algorithm can be implemented using the FACF in simulations. Finally, we conclude this work with a discussion in Sec.~\ref{sec:END}.

\section{Foundations of the noise-cancellation method} \label{sec:NCA}

Let us first reiterate the idea of the NC algorithm as it was introduced and applied in the recent works Refs.~\cite{Mandal2019, Rusch2024}, dealing with efficiently computing the MSD and VACF of soft-matter systems. The central idea behind the NC algorithm utilizes a decomposition of the total displacement $\Delta \vec{R}(t)$ of a tagged particle over the time $ t $ (where we set the initial time $ t_0 = 0 $ throughout this paper) into two physically distinct components~\cite{Mandal2019, Rusch2024} 
\begin{equation}
    \Delta \vec{R}(t) = \delta \vec{R}(t) + \Delta \vec{R}_0(t), \label{eq:LININT}
\end{equation}
where $\delta \vec{R}(t)$ represents the contribution from interparticle interactions or external forces, while $\Delta \vec{R}_0(t)$ denotes the displacement corresponding to the hypothetical motion of a free Brownian particle. Such a decomposition always exists for Brownian systems due to the linearity of the time integration. For its application in simulations, it is usually implemented by performing two simulations with identical noise histories --- one that includes particle-particle interactions and external potentials and one that omits them~\cite{Rusch2024}.

Applying the decomposition Eq.~\eqref{eq:LININT} twice to the definition of the MSD (once to expand it in multiple terms and once to substitute $ \Delta \vec{R}_0(t) $ in the cross-correlation term), the MSD can be expressed as
\begin{equation}
    \left\langle \Delta \vec{R}^2(t) \right\rangle = 2 d D_0 t - \left\langle \delta \vec{R}^2(t) \right\rangle + 2 \left\langle \Delta \vec{R}(t) \cdot \delta \vec{R}(t) \right\rangle, \label{eq:MSD}
\end{equation}
where $D_0$ is the short-time diffusion coefficient of the particle. Here, $ \langle \cdot \rangle $ denotes an ensemble average, which in simulations is typically estimated using a sliding-time average under the assumption of ergodicity~\cite{Rusch2024}. Such an average usually takes the form $ \langle \Delta f(\tau)\, \Delta g(\tau) \rangle = \frac{1}{M} \sum_i \left[f(t_i+\tau) - f(t_i)\right]\left[g(t_i+\tau) - g(t_i)\right] $, where the $ t_i $ are $ M $ predefined sampling times along the tagged-particle trajectory, and $ f $ and $ g $ are arbitrary observables. The first term of Eq.~\eqref{eq:MSD} reflects the exact result for a free Brownian particle, $ \langle \Delta \vec{R}_0^2(t) \rangle = 2dD_0t $. By incorporating this known relation in the MSD expression, an explicit dependence on stochastic noise is eliminated, thereby significantly reducing statistical uncertainty, especially at long times.

Up to this point, the NC procedure is comparable to the considerations made by Frenkel~\cite{Frenkel1987}, who used a similar strategy to decompose the particle velocity instead of the displacement in his computation scheme to calculate velocity correlations. To further suppress noise, Mandal~\emph{et al.}~\cite{Mandal2019} proposed a crucial simplification: the cross-correlation term $\left\langle \Delta \vec{R}(t) \cdot \delta \vec{R}(t) \right\rangle$ and its time derivatives can be neglected. This leads to a simplified and computationally efficient approximation of the MSD
\begin{equation}
     \left\langle \Delta \vec{R}^2(t) \right\rangle \approx 2 d D_0 t - \left\langle \delta \vec{R}^2(t) \right\rangle. \label{eq:NCF}
\end{equation}
In this NC approximation, the thermal noise enters only implicitly via $\delta \vec{R}(t)$, improving the signal-to-noise ratio, particularly in dilute or weakly interacting systems~\cite{Rusch2024}. The effectiveness of this approximation depends on the relative magnitude of the neglected cross-term $\left\langle \Delta \vec{R}(t) \cdot \delta \vec{R}(t) \right\rangle$ compared to $\left\langle \delta \vec{R}^2(t) \right\rangle$. Previous studies have empirically shown that the cross-correlations are orders of magnitude smaller in dilute suspensions~\cite{Mandal2019, Rusch2024}, but a general theoretical criterion for the validity of neglecting the cross-correlations has so far been lacking.

To extract the VACF $Z(t) $ from Eq.~\eqref{eq:NCF}, the general relation
\begin{equation}
    Z(t) = \frac{1}{2d} \frac{\textrm{d}^2}{\textrm{d}t^2} \left\langle \Delta \vec{R}^2(t) \right\rangle, \label{eq:DEFVACF}
\end{equation}
can be applied. This expression will act as the definition for the VACF throughout this work [note that, for Brownian particles, the velocity is not a well-defined observable, but the VACF can always be defined by Eq.~\eqref{eq:DEFVACF}]. Neglecting the second time derivative of the cross-correlations, the NC approximation of the VACF then reads~\cite{Rusch2024}
\begin{equation}
    Z(t) \approx - \frac{1}{2d} \frac{\textrm{d}^2}{\textrm{d}t^2} \left\langle \delta \vec{R}^2(t) \right\rangle, \label{eq:VACFNC}
\end{equation}
for $ t > 0 $. This approximation forms the basis of the NC algorithm for the VACF: by isolating the interaction-induced displacement $\delta \vec{R}(t)$ and applying a second time derivative to its mean square, one can reconstruct the VACF while minimizing thermal noise. As demonstrated in Ref.~\cite{Mandal2019}, this approach leads to a significant improvement in resolving the long-time decay of the VACF in interacting Brownian particles.

\section{General relation between the VACF and the FACF} \label{sec:Gen}

In this section, we aim to establish a rigorous theoretical foundation for the NC algorithm by identifying the conditions under which the cross-correlation term in Eq.~\eqref{eq:DEFVACF} can be safely neglected in Brownian systems. To this end, we derive a general relation between the VACF, the FACF, and the associated cross-correlation contributions. This relation not only clarifies the physical origin of the NC approximation but also serves as the analytical basis for evaluating its accuracy.

In the following, we restrict ourselves to overdamped Brownian systems, which can be described by an equation of motion of the form
\begin{equation}
    \dot{\vec{R}}(t) = \mu \, \vec{F}(t, \vec{r}(t)) + \boldsymbol{\eta}(t). \label{eq:LE}
\end{equation}
Here, $ \mu = D_0 / k_B T $ is the mobility, $ \boldsymbol{\eta}(t) $ is the translational noise defined by $ \langle \boldsymbol{\eta}(t) \rangle = 0 $ and $ \langle \boldsymbol{\eta}(t) \otimes \boldsymbol{\eta}(t') \rangle = 2D_0 \mathds{1}\,\delta(t - t') $ (where the average is taken over an infinite number of independent realizations of the noise~\cite{Zwanzig2001}). The vector $ \vec{r} $ summarizes all translational degrees of freedom of the system, $ \otimes $ denotes the dyadic product, and $ \mathds{1} $ is the unit matrix. The force $ \vec{F}(t, \vec{r}) $ may include conservative, non-conservative, or effective contributions; the latter two can drive the \qq{tagged} degrees of freedom $ \vec{R}(t) $ of the system out of equilibrium.

Throughout this work, the overdamped Langevin equation~\eqref{eq:LE} can describe either a single Brownian particle ($ \vec{R}(t) = \vec{r}(t) $) subject to an external force (as in Ref.~\cite{Rusch2024}) or an arbitrarily chosen tagged particle in a colloidal suspension (as in Ref.~\cite{Mandal2019}). For the latter, $ \vec{F}(t, \vec{r}) $ generally contains particle-particle interactions that depend on the positions $ \vec{r}(t) = (\vec{r}_1(t), \ldots, \vec{r}_N(t)) $ of all particles at time $t$, and $ \vec{R}(t) = \vec{r}_i(t) $ denotes the position of the chosen tagged particle $i$. In this case, $ \vec{F}(t) $ should always be interpreted as $ \vec{F}(t, \vec{r}(t)) $, where we write only the explicit $ t $ dependence for simplicity.

Finally, we only look at systems that are stationary in the sense of $ {\langle \dot{\vec{R}}(t) \cdot \dot{\vec{R}}(t')\rangle = \langle \dot{\vec{R}}(\tau) \cdot \dot{\vec{R}}(0)\rangle = \langle \dot{\vec{R}}(-\tau) \cdot \dot{\vec{R}}(0)\rangle} $ and $ {\langle \vec{F}(t) \cdot \vec{F}(t')\rangle = \langle \vec{F}(\tau) \cdot \vec{F}(0)\rangle = \langle \vec{F}(-\tau) \cdot \vec{F}(0)\rangle} $, where $ \tau = t - t' $ is the lag time. Here, $ \langle \cdot \rangle $ is determined by the noise average above, and describes the mean over an ensemble, characterized by different realizations of the thermal noise $ \boldsymbol{\eta}(t) $ and initial conditions drawn from the probability density that describes the assumed stationary state. However, we do not assume thermal equilibrium (in particular, detailed balance) yet.

We can connect the NC relations~\eqref{eq:MSD}, \eqref{eq:NCF}, and \eqref{eq:VACFNC} with the terms of the Langevin equation. Integrating Eq.~\eqref{eq:LE} with respect to the time, we identify the contributions $ \delta \vec{R}(t) $ and $ \Delta \vec{R}_0(t) $ as defined in Sec.~\ref{sec:NCA} via 
\begin{align}
    \delta \vec{R}(t) & = \mu \int_0^t \vec{F}(s) \, \textrm{d} s \label{eq:deltaR},\\
    \Delta \vec{R}_0(t) & = \int_0^t \boldsymbol{\eta}(s) \, \textrm{d} s. \label{eq:DeltaR}
\end{align}
By our assumptions of stationarity, we infer that
\begin{align}
    \left\langle \delta \vec{R}^2(t) \right\rangle & = \mu^2 \int_0^t \int_0^t \langle \vec{F}(s) \cdot \vec{F}(s')\rangle \, \textrm{d}s \textrm{d}s' \nonumber \\ & = 2 \mu^2 \int_0^t \int_0^s \langle \vec{F}(s') \cdot \vec{F}(0)\rangle \, \textrm{d}s' \textrm{d}s, \label{eq:REWRITE}
\end{align}
which has the same form as the well-established relation between total displacement $ \langle \Delta \vec{R}^2(t) \rangle $ and the velocity correlations that is the basis of the definition of the VACF given by Eq.~\eqref{eq:DEFVACF}. With this, we can now calculate the second derivative with respect to the time of Eq.~\eqref{eq:MSD}, and obtain the VACF
\begin{equation}
    Z(t) = -\frac{\mu^2}{d} \left\langle \vec{F}(t) \cdot \vec{F}(0) \right\rangle + \frac{1}{d} \frac{\textrm{d}^2}{\textrm{d}t^2}  \left\langle \Delta \vec{R}(t) \cdot \delta \vec{R}(t) \right\rangle, \label{eq:VACFGENERALES}
\end{equation}
for $ t > 0$ from Eq.~\eqref{eq:DEFVACF}. Here, the bare diffusion term of the MSD of the order $O(t)$ drops out when calculating the derivatives while formally contributing a Dirac delta peak $ 2dD_0\delta(t) $ at $ t = 0 $~\cite{Frenkel1987, Mandal2019}. Our derivation thus far only required the assumption of stationarity (in particular, that the correlation functions are only even functions of the lag time) to rewrite the mean square of the deterministic displacement $ \delta \vec{R}(t) $ via Eq.~\eqref{eq:REWRITE}. Consequently, it holds for both Brownian equilibrium and nonequilibrium systems that can be described by the overdamped Langevin equation Eq.~\eqref{eq:LE}. 

Note, however, that the definition of the noise term of Eq.~\eqref{eq:LE} implicitly assumes that the surrounding fluid obeys the fluctuation-dissipation theorem, which is therefore in equilibrium. In other words, we assume that the forces $\vec{F}(t)$ do not significantly perturb the fluid, either because the forces are small, or because the fluid relaxes much faster than the Brownian particle evolves. Thus, when we refer to a \qq{Brownian non-equilibrium system}, we mean that the \emph{tagged particle} is out of equilibrium (i.\@e.\@, its probability density $P(\vec{R},t)$ does not follow the equilibrium distribution), while the \emph{background fluid} is still treated as an equilibrium bath. This approximation is widely used in simulations of colloidal non-equilibrium systems, including active particles~\cite{Bechinger2016} and externally driven particles~\cite{Isele2023, Foss2000}.

Equation~\eqref{eq:VACFGENERALES}, which shows that the VACF is given by the FACF and a term depending on the time derivatives of the cross-correlations, is the foundation for our work. As we will discuss in the next sections, a direct consequence of Eq.~\eqref{eq:VACFGENERALES} is that the second time derivative of the cross-correlations always vanishes in the equilibrium. Additionally, it can be used to assess the impact of the cross-correlations on the VACF of Brownian nonequilibrium systems, as we will argue later.

\section{Equilibrium} \label{sec:EQ}

For a homogeneous system in equilibrium, the configuration of the system can be prescribed by the collection of positions $\vec{r} = (\vec{r}_1,\ldots, \vec{r}_N)$. The particles are expected to interact with each other via a time independent many-body potential $U=U(\vec{r})$ and the corresponding equilibrium distribution can be described by the canonical probability distribution $P_{\textrm{eq}}(\vec{r}) \propto \exp[ - U(\vec{r})/k_B T]$. Also, the corresponding average over the noise and the probability density of the initial values $ \langle \cdot \rangle_{\textrm{eq}} $ is equivalent to the canonical ensemble average when applied to observables that only depend on the particle configuration. For simplicity, we consider the first particle as the \qq{tagged particle}, $\vec{R}(t) = \vec{r}_1(t)$, the dynamics of which we monitor and describe with the overdamped Langevin equation~\eqref{eq:LE}. Thus, the MSD can then be computed by
\begin{align}
    \left\langle \Delta \vec{R}^2(t - t')  \right\rangle_{\textrm{eq}} =    \int\!\!\int \textrm{d}\vec{r} \, \textrm{d} \vec{r}' \, ( \vec{r}_1 - \vec{r}_1' )^2 \, \mathds{P}(\vec{r},t|\vec{r}', t') P_{\textrm{eq}}(\vec{r}'), \label{eq:MSDEQ}
\end{align}
where $\mathds{P} =  \mathds{P}(\vec{r},t\mid \vec{r}', t') $ corresponds to the conditional probability to find the system in configuration $\vec{r} = (\vec{r}_1,\ldots, \vec{r}_N)$ at time $ t $ given it was in configuration $\vec{r}' = (\vec{r}_1',\ldots, \vec{r}_N')$ at earlier time $ t' $. The probability density $\mathds{P}$, also referred to as the propagator, solves the $ N $-particle Smoluchowski equation
\begin{equation}
    \left[\partial_t - \hat{\Omega}_{\vec{r}} \right] \, \mathds{P}(\vec{r},t \mid \vec{r}', t') = 0, \label{eq:Smol}
\end{equation}
which directly results from our overdamped Langevin equations describing the Brownian system. 
Here, $ \hat{\Omega}_{\vec{r}} $ is the $ N $-particle Smoluchowski operator acting on functions ${A=A(\vec{r})}$ of the final configuration as
\begin{equation}
    \hat{\Omega}_{\vec{r}} A  = \sum_i \left[D_0 \,\nabla^2_{\vec{r}_i} A  + \mu \nabla_{\vec{r}_i} \cdot ( A  \nabla_{\vec{r}_i} U )  \right]. 
\end{equation}
Similarly, the propagator fulfils the corresponding Kolmogorov backward equation
\begin{equation}
     \left[\partial_{t'} + \hat{\Omega}_{\vec{r'}}^{\dagger} \right] \, \mathds{P}(\vec{r} ,t \mid \vec{r}', t') = 0, \label{eq:KBE}
\end{equation}
with the adjoint operator $ \hat{\Omega}_{\vec{r}'}^{\dagger} $ acting on functions of the initial configuration $B=B(\vec{r}')$ as 
\begin{equation}
    \hat{\Omega}_{\vec{r}'}^{\dagger} B  = \sum_i \left[D_0\,\nabla^2_{\vec{r}'_i} B  - \mu \,\nabla_{\vec{r}_i'} U \cdot \nabla_{\vec{r}_i'} B \right]. \label{eq:BACKOP}
\end{equation}
Calculating the derivatives of Eq.~\eqref{eq:MSDEQ} with respect to $ t $ and $ t' $, inserting the Smoluchowski equation and the Kolmogorov backward equation, and using the identity $ {P_{\textrm{eq}}( \vec{r}')  \hat{\Omega}_{\vec{r}'}^{\dagger} (\ldots) = \hat{\Omega}_{\vec{r}'} [ (\ldots) P_{\textrm{eq}}(\vec{r}')]} $~\cite{Jones2003}, we find
\begin{equation}\label{eq:VACFEQ}
    Z_{\textrm{eq}}(t-t') = -\frac{\mu^2}{d}  \left\langle \vec{F}(t) \cdot \vec{F}(t') \right\rangle_{\textrm{eq}}, 
\end{equation}
where  $\vec{F}(t) \equiv \vec{F}(\vec{r}(t)) = - (\nabla_{\vec{R}} U)(\vec{r}(t))$ denotes the force acting on the tagged particle at time $t$. A detailed derivation is given in Appendix~\ref{app:eq}. This established identity, which holds for arbitrary Brownian systems in equilibrium, was first derived by Hanna~\emph{et al.}\@~\cite{Hanna1981} to analytically obtain the VACF of hard spheres. A more general expression that also holds for odd-diffusive systems (i.\@e.\@, systems where $ D_0 $ is given by an asymmetric tensor) was recently calculated in Ref.~\cite{Kalz2024}. See also Refs.~\cite{Cengiov2019, Sharma2016} for an application of this identity in the context of linear response and active matter.

Strikingly, by comparing Eq.~\eqref{eq:VACFEQ} with our general relation Eq.~\eqref{eq:VACFGENERALES}, we find
\begin{equation}
    \frac{1}{d} \frac{\textrm{d}^2}{\textrm{d}t^2}  \left\langle \Delta \vec{R}(t) \cdot \delta \vec{R}(t) \right\rangle_{\textrm{eq}} \equiv 0, \label{eq:CCAREZERO}
\end{equation}
i.\@e.\@, the second time derivative of the cross-correlation term always vanishes in thermal equilibrium. Hence, the NC algorithm and its Eq.~\eqref{eq:VACFNC} are not approximations but exact expressions for Brownian equilibrium systems, which is one of the key findings of this work.

The identity between the VACF and the negative FACF in the equilibrium gives rise to an alternative approach of implementing the NC algorithm in simulations: Instead of computing the MSD via Eq.~\eqref{eq:NCF} and calculating the time derivatives numerically to obtain the VACF (as done previously~\cite{Mandal2019, Rusch2024}), the FACF can be calculated via the forces applied to evolve the motion of the particles to compute the velocity correlations directly. This FACF-based approach inherently benefits from noise cancellation, since in Brownian systems with short-range interactions, forces vanish when particles are not interacting and only become significant during collisions.

The alternative formulation based on the FACF highlights the connection of the NC algorithm to the computation scheme of \AA{}kesson~\emph{et al.}\@~\cite{Akesson1985}, which utilizes the FACF to obtain the long-time diffusion coefficient for Brownian systems using similar principles.The method of \AA{}kesson~\emph{et al.}\@ has been applied in numerous simulation studies~\cite{Hartl1992, Heyes1994, Heyes1995, Heyes1996, Heyes2000, Branka2005} to enhance convergence by reducing statistical noise, and a similar relation has independently been derived for systems including hydrodynamic interactions by Fixman~\cite{Fixman1981}.
Notably, the findings of Rusch~\emph{et al.}\@~\cite{Rusch2024}, which demonstrate improved precision of the NC algorithm in weakly interacting systems, align closely with those of Hartl~\emph{et al.}\@~\cite{Hartl1992}, who argued that the FACF-based approach to numerically compute long-time diffusion coefficients of \AA{}kesson~\emph{et al.}\@ performs best when interparticle coupling forces are small. 

Beyond this, Eq.~\eqref{eq:VACFEQ} also gives rise to further connections to other computation schemes utilized in the literature. For instance, the NC algorithm is in line with works such as Ref.~\cite{Heyes2002}, where it is found that the FACF can be used to improve the statistical resolution when computing the series expansion of the VACF of Newtonian particles, or Ref.~\cite{Rastogi1996}, where the MSD of a Brownian particle is also reduced to a FACF-based term. See also Ref.~\cite{Liu2003}, where a similar relation was applied for the diffusion coefficient of Brownian particles interacting via hydrodynamic interactions.

Taken together, these connections demonstrate that the NC algorithm is not only exact in equilibrium but also naturally aligned with a broad class of established force-based computational techniques. However, to the best of our knowledge, the FACF has not been suggested to efficiently study the long-time tails of the VACF of Brownian systems before. In the next section, we apply the force-based computational techniques to specific equilibrium model systems to assess the practical accuracy of the NC algorithm. Our analysis will emphasize the central role of the VACF-FACF relation, illustrating how FACF-based approaches serve as a robust alternative implementation of the NC algorithm, especially effective in capturing long-time tails of the VACF, as also reported in Ref.~\cite{Mandal2019}.

\subsection{VACF long-time tails for soft spheres}

First, we focus on the applicability of the FACF-based approach of obtaining noise-suppressed data for the VACF in equilibrium many-body simulations: One of the biggest achievements of the NC algorithm so far is the resolution of the long-time tails of the VACF in dilute Brownian hard-sphere systems~\cite{Mandal2019} within a reasonable sample size. A common approach to approximate hard spheres in many-body simulations is using repulsive soft spheres modeled via the purely-repulsive Weeks-Chandler-Andersen (WCA) potential, given by
\begin{align}
     V(r) = \left\{ 
     \begin{array}{ll}
     4 \varepsilon \left[ \left( \sigma/r \right)^{12} - \left( \sigma/r \right)^6 \right] + \varepsilon ,& r \le 2^{1/6} \sigma \\
     0, & r > 2^{1/6} \sigma.
     \end{array}
     \right. \label{eq:WCA}
\end{align}
Here, $ \sigma $ is the diameter of the particles and $ \varepsilon $ the interaction strength (which we set $ \varepsilon = 1\,k_BT $ throughout this work for simplicity). For such soft-sphere simulations, the forces $ ( \vec{F}_1(t), \ldots, \vec{F}_N(t)) $ acting on the particles are easily accessible through the negative gradient of the pair potential. In fact, the forces have to be evaluated at each time step to iteratively solve the particle movements in simulations either way. Hence, the FACF can easily be determined during the simulations to obtain the VACF \qq{on-the-fly}. 

\begin{figure}[ht!]
\includegraphics[width=0.49\textwidth]{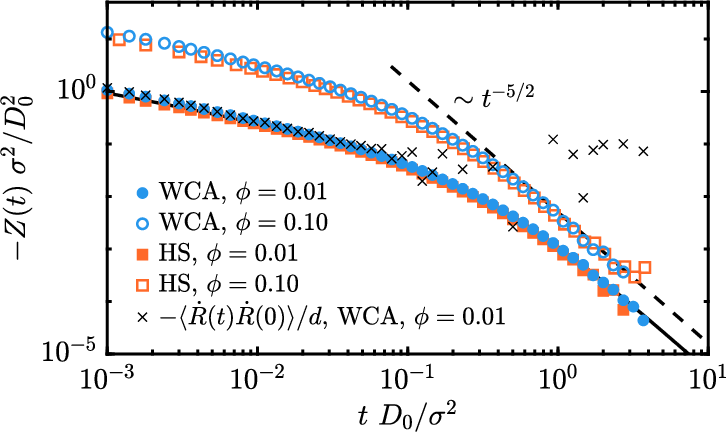}
\caption{Velocity autocorrelation functions (VACF) of three-dimensional Brownian particles. Solid line is the analytical hard-sphere result for small volume fractions $ \phi $. Dashed line acts as a guide for the eye for the long-time tail corresponding to the exponent $ - 5/2 $. Circles corresponds to the results of particles that interact via the WCA potential. Squares depict the results of the hard-sphere simulations. Crosses are calculated by directly correlating the displacements per simulation time step of the WCA particles. All other data is calculated using the force autocorrelation function (FACF). See Appendix~\ref{app:WCA} and Appendix~\ref{app:HS} for the simulation parameters.}
\label{fig:1}
\end{figure}

To show the validity of the FACF-based method, we perform BD simulations (i.\@e.\@, we numerically integrate the overdamped Langevin equations) of colloidal particles that interact via the WCA potential for the volume fractions of $ \phi = 0.01 $ and $ \phi=0.10$ (see Appendix~\ref{app:WCA} for details regarding the utilized algorithm and the simulation parameters). As the system is purely repulsive, we can compare the results with the exact theoretical solution for the VACF of hard spheres~\cite{Hanna1981, Mandal2019}
\begin{align}
    Z_\textrm{eq}(t)  =& -8\phi \frac{D_0^2}{\sigma^2} \left\{ \sqrt{\frac{\tau_D}{2\pi t}} \right. \nonumber\\
    & - \cos(4t/\tau_D) \left[1-2\textrm{S}(\sqrt{8t / \pi \tau_D})\right]  \nonumber\\ & + \left. \sin(4t/\tau_D) \left[1-2\textrm{C}(\sqrt{8t/\pi \tau_D}) \right] \right\} + O(\phi^2), \label{eq:ZEQ}
\end{align}
for intermediate and long times, where $ \tau_D = \sigma^2 / D_0$ is the Brownian time, and $\textrm{S}(\cdot)$, $ \textrm{C}(\cdot) $ denote the Fresnel integrals~\cite{Hanna1981, Mandal2019}.

In Fig.\ref{fig:1}, the circles indicate the VACF obtained using the FACF. For low volume fractions, the data closely match the analytical prediction (line), while successfully reproducing the long-time tail with the characteristic $t^{-5/2}$ decay at $\phi=0.10$. These results are consistent with those of Mandal~\emph{et al.}\@~\cite{Mandal2019}, confirming that the FACF-based NC implementation is a robust and accurate method. For comparison, the data marked with crosses show results from the same simulations but using (approximate) particle velocities $ {\dot{\vec{R}}(t) \approx [\vec{R}(t+\Delta t) - \vec{R}(t)] / \Delta t} $ to calculate the VACF directly without noise suppression. This approach fails to resolve the long-time tail, highlighting the clear advantage of the FACF-based method in suppressing noise and enhancing statistical resolution.

\subsection{VACF long-time tails for hard spheres}

While the computation of the FACF is straightforward in simulations of soft particles, where the forces $(\vec{F}_1(t), \ldots, \vec{F}_N(t) )$ are readily available from the pair potential, it remains an open question whether a similar FACF-based implementation of the NC algorithm can be extended to simulations of hard spheres. In those systems, no explicit analytic expression for the interaction forces exists. To address this, we present a proof of concept that an effective FACF can still be used to apply the NC scheme to hard-sphere systems. We do this by employing the \qq{potential-free} algorithm of Heyes and Melrose~\cite{Heyes1993}. This simulation method, which proved to be a viable option in various BD simulation-based studies~\cite{Zia2010, Zia2012, Rintoul1996}, was successfully applied to approximate interaction forces in hard-sphere systems to compute stress correlation functions in good agreement with analytic predictions~\cite{Sunol2023, Marenne2017, Foss2000}.

In the \qq{potential-free} BD algorithm, the particles are first displaced to address the thermal fluctuations. Afterwards, it is pairwise checked if two particles overlap. If this is the case, the centers of the overlapping particles are moved by a displacement $ \Delta \vec{r}_i^{\textrm{HS}} $ so that the hard spheres are in touch. To obtain an approximation for the (effective) forces, we use the method presented in Refs.~\cite{Sunol2023, Marenne2017, Foss2000}. Here, $\vec{F}_i(t)$ is estimated by the Stokesian force required to move particle $i$ by $\Delta \vec{r}_i^{\textrm{HS}}$ over a simulation time step of duration $\Delta t$. This corresponds to the necessary momentum that must be transferred to the particle $ i $ per time step to achieve the prescribed displacement $ \Delta \vec{r}_i^{\textrm{HS}} $. Hence, we have
\begin{equation}
    \mu\,\vec{F}_i(t) \approx \frac{ \Delta \vec{r}_i^{\textrm{HS}}}{\Delta t}, \label{eq:HSForce}
\end{equation}
which directly connects $ \Delta \vec{r}_i^{\textrm{HS}} $ to the deterministic force term of the overdamped Langevin equation~\eqref{eq:LE}.

While this method is less accurate than event-driven Brownian dynamics (EDBD) techniques, such as those used in Ref.~\cite{Mandal2019}, due to its limited resolution of multi-particle collisions, it has proven sufficiently reliable for computing stress correlations even at moderate to high volume fractions (up to $\phi=0.50$)~\cite{Sunol2023, Marenne2017, Foss2000}. As such, it is expected to be more than adequate for our proof of concept of the FACF-based NC scheme for hard spheres in the dilute regime. Additional details on the conducted simulations and the corresponding parameters are provided in Appendix~\ref{app:HS}.

The results for $Z_{\textrm{eq}}(t)$ obtained via the effective FACF for the hard-sphere systems (with $ \phi = 0.01 $ and $ \phi = 0.10 $) are marked by the squares in Fig.~\ref{fig:1}. The results follow the theoretical prediction and the established behavior found in Ref.~\cite{Mandal2019} sufficiently. This demonstrates that an effective FACF constructed from physically motivated approximations of the forces $(\vec{F}_1(t), \ldots, \vec{F}_N(t) ) $ can be successfully employed to compute the VACF via the NC algorithm, even in the absence of a well-defined pair potential. Given this success with a relatively simple algorithm, we anticipate that even better accuracy can be achieved with more sophisticated simulation strategies.

\subsection{Harmonic potential}

As a case study for a single particle in an external potential, we revisit a Brownian particle in a harmonic trap [i.\@e.\@, $ {\vec{F}(t) = -k \vec{R}(t)} $], which was already analyzed in the context of the NC algorithm for $ d = 1 $ in Ref.~\cite{Rusch2024}. For this setup, Eq.~\eqref{eq:VACFEQ} can be validated analytically. Since the corresponding equation of motion is an Ornstein-Uhlenbeck process, one evaluates~\cite{Gao2006, Goerlich2021}
\begin{equation}
    \vec{R}(t) = \vec{R}(0)\,\textrm{e}^{-\mu k t} + \int_0^t \textrm{e}^{-\mu k (t-s) }\, \boldsymbol{\eta}(s) \, \textrm{d}s, \label{eq:SOLHARM}
\end{equation}
which directly results in
\begin{equation}
    \left\langle \vec{F}(t) \cdot \vec{F}(0)\right\rangle_{\textrm{eq}} = \frac{ kD_0 d}{\mu} \exp(-\mu k t), 
 \label{eq:FNCFHP}
\end{equation}
for $ t > 0 $, if the equilibrium mean-squared position $ {\langle \vec{R}^2\rangle_{\textrm{eq}} =  D_0 d/ \mu k} $ is inserted. 

Next, to derive the VACF of a particle in a harmonic potential, we can insert the general Langevin equation~\eqref{eq:LE} in the standard definition $ Z(t) = \langle \dot{\vec{R}}(t) \cdot \dot{\vec{R}}(0) \rangle / d $. By simple algebraic manipulations, we find
\begin{equation}
    Z(t) = \frac{\mu^2}{d} \left\langle \vec{F}(t) \cdot\vec{F}(0) \right\rangle + \frac{\mu}{d} \left\langle \vec{F}(t+\tau_0) \cdot \boldsymbol{\eta}(\tau_0)\right\rangle ,
\label{eq:VACFNOISE}
\end{equation}
for $ t > 0 $ and arbitrary $ \tau_0 > 0 $, where we used that $ {\langle \vec{F}(\tau_0) \cdot \boldsymbol{\eta}(\tau_0+t) \rangle} $ must vanish for $ t > 0 $ as the noise is independent of the drift term. Since $ Z(t) $ and the FACF are independent of the time origin by our assumptions, $ \left\langle \vec{F}(t+\tau_0) \cdot \boldsymbol{\eta}(\tau_0)\right\rangle $ must be independent of $\tau_0$. However, as correlations with $ \boldsymbol{\eta}(0) $ in the Langevin formalism can be misinterpreted when integrating with respect to the time (because delta functions are shifted to the integral boundaries), we here refer this quantity to a fixed $ \tau_0$.

Applying the solution~\eqref{eq:SOLHARM} for $ \vec{R}(t) $ to Eq.~\eqref{eq:VACFNOISE} and inserting the equilibrium FACF, we finally obtain
\begin{equation}
    Z_{\textrm{eq}}(t) = - \mu k D_0\exp(-\mu k t).
\end{equation}
Comparing this result with the FACF~\eqref{eq:FNCFHP} confirms that Eq.~\eqref{eq:VACFEQ} holds for this example system and that the NC algorithm is an exact expression after the system is fully equilibrated [while equilibrating, $ {\langle \vec{R}^2(t) \rangle \neq \langle \vec{R}^2\rangle_{\textrm{eq}}} $ and the NC algorithm cannot be applied].

Indeed, the NC algorithm was already proven to be exact for a one-dimensional Brownian particle in a harmonic potential by Rusch~\emph{et al.}\@~\cite{Rusch2024}. While deriving their results, they proved an additional property of the cross-correlations, which we want to highlight here since it holds important insight into the accuracy of the NC computation scheme: For arbitrary discretizations of the underlying Langevin equations with a step size $ \Delta t $, the cross-correlation term for the harmonic potential yields~\cite{Rusch2024}
\begin{equation}
     \left\langle \Delta \vec{R}(t) \cdot \delta \vec{R}(t) \right\rangle_{\textrm{eq}} =  D_0d \Delta t \left[ 
1-\exp(-\mu k t) \right] + O(\Delta t^2),
\end{equation}
which means it is finite and of order $O(\Delta t) $. Hence, our result that the NC algorithm is exact in equilibrium is formally only true for the continuous case with $ \Delta t \rightarrow 0 $. If the VACF is computed with simulations, there will be an error arising from the discretizations of the dynamics (which is not surprising for computer simulations). However, this error can be made arbitrarily small by adjusting the step size $ \Delta t $ and we expect it to be similar in order compared to other uncertainties introduced by discretizing the dynamics. 

\subsection{Force-noise correlations}

Another interesting relation can be obtained from equating Eqs.~\eqref{eq:VACFGENERALES} and \eqref{eq:VACFNOISE} by solving for the force-noise correlations: We find 
\begin{align}
    \left\langle \vec{F}(t+\tau_0) \cdot \boldsymbol{\eta}(\tau_0)\right\rangle  =& -2 \mu \left\langle \vec{F}(t) \cdot \vec{F}(0)\right\rangle \nonumber \\ & + \frac{1}{\mu} \frac{\textrm{d}^2}{\textrm{d}t^2}  \left\langle \Delta \vec{R}(t) \cdot \delta \vec{R}(t) \right\rangle, \label{eq:FNCF}
\end{align}
for $ t > 0 $, which shows that force-noise correlations for an arbitrary $ \vec{F}(t) $ are always given by the FACF and the second time derivative of the cross-correlation term.

In thermal equilibrium, Eq.~\eqref{eq:FNCF} simplifies to the nontrivial result
\begin{equation}
    \left\langle \vec{F}(t+\tau_0) \cdot \boldsymbol{\eta}(\tau_0)\right\rangle_{\textrm{eq}} = -2 \mu \left\langle \vec{F}(t) \cdot \vec{F}(0)\right\rangle_{\textrm{eq}}, \label{eq:NFCF}
\end{equation}
with $ t > 0 $ [see Eq.~\eqref{eq:CCAREZERO}]. Thus, the noise-force correlations always follow the negative FACF, completely independent of the actual pair interactions between the particles and the form of the external potential. This fundamental relation was first derived in Ref.~\cite{Akesson1985} in the context of Brownian particles interacting through electrostatic potentials, and later confirmed for Brownian systems with hydrodynamic interactions~\cite{Liu2003}. Therefore, Eq.~\eqref{eq:FNCF} (which also holds in stationary nonequilibrium as Eq.~\eqref{eq:VACFGENERALES} only requires stationarity) expands on the equilibrium results of Ref.~\cite{Akesson1985}.

\begin{figure}[ht!]
\includegraphics[width=0.49\textwidth]{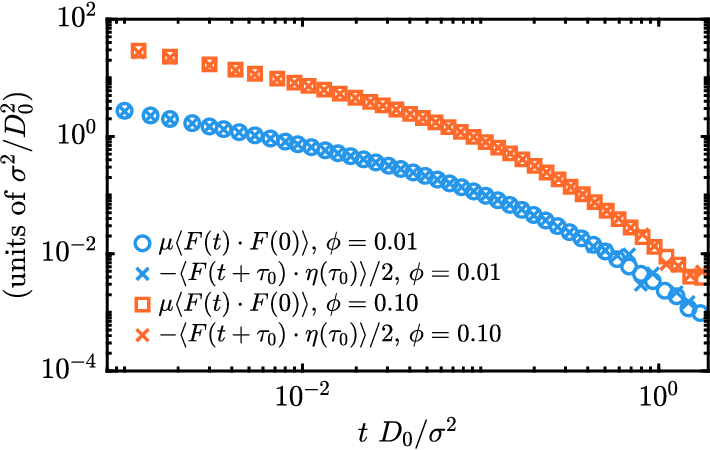}
\caption{Comparison between the force-noise correlations and the force autocorrelation functions for Brownian hard-sphere suspensions at different volume fractions. See Appendix~\ref{app:HS} for the corresponding simulation parameters.}
\label{fig:2}
\end{figure}

To test whether Eq.~\eqref{eq:NFCF} is valid for hard spheres (it has only been studied for continuous interactions so far~\cite{Akesson1985, Liu2003}), we perform BD simulations and compare its left-hand side with the effective force autocorrelation function (FACF). These results are shown in Fig.~\ref{fig:2} and we use the same simulation parameter as for the calculation of the VACF of hard spheres. Keep in mind that force-noise correlations depend explicitly on the noise, which causes more fluctuations compared to the effective FACF. Strikingly, both correlation functions match closely. Similar results are also seen in BD simulations using soft WCA particles (not shown here to keep things concise). Our numerical evidence supports the analytical predictions: in systems driven by thermal fluctuations, where particle displacements and interaction effects are uncorrelated, the correlation between the deterministic drift and the random force in the Langevin equation follows the autocorrelation of the deterministic drift itself.

Note that we did not define Eqs.~\eqref{eq:FNCF} and \eqref{eq:NFCF} for $t = 0$ in this section. In general, correlations between particle positions at time $t$ and the noise evaluated at the same time can depend on the chosen interpretation of the stochastic calculus. Therefore, special care is required when extending our results to $t = 0$.

\subsection{Single-file diffusion}

To further demonstrate the versatility of the FACF-based NC algorithm, we apply it to an example system that has not previously been analyzed with the original NC scheme, namely many-body hard-sphere systems in $ d = 1 $. For this purpose, we employ the potential-free BD algorithm described earlier. See Appendix~\ref{app:HS} for details and the simulation parameters.

\begin{figure}[ht!]
\includegraphics[width=0.49\textwidth]{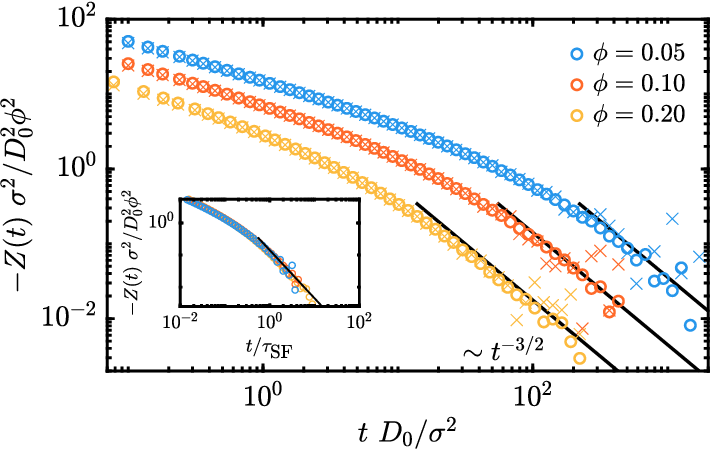}
\caption{Velocity autocorrelation function (VACF) for one-dimensional Brownian hard spheres at different densities $ \phi $. Circles show the results obtained via the effective force autocorrelation function (FACF). Crosses correspond to data obtained with a standard (noise suppressing) algorithm~\cite{Frenkel1987}. Solid lines are the predicted long-time behaviors [Eq.~\eqref{eq:VACFSFD}]. Inset: Curves for different $ \phi $ plotted with respect to the single-file time scale $ \tau_{\textrm{SF}} $. See Appendix~\ref{app:HS} for the corresponding simulation parameters.}
\label{fig:3}
\end{figure}

One of the most well-known consequences of particle-number conservation in one-dimensional systems is single-file diffusion~\cite{Marchesoni2006, Taloni2006, Flomenbom2008, Taloni2008, Taloni2017, Wei2000, Kollmann2003, Lutz2004}. When particles are confined to a one-dimensional geometry and cannot overtake each other, the mean-squared displacement (MSD) becomes subdiffusive, scaling as $t^{1/2}$ in the long-time regime. Hence, for the VACF, a long-time tail with exponent $ t^{-3/2} $ emerges. While single-file diffusion can easily be observed for intermediate particle densities, it can be hard to resolve in the dilute regime using Langevin-type equations of motion. This is because the time scale $ \tau_{\textrm{SF}} $ on which the interactions with the neighboring particles become noticeable is given by ${\tau_{\textrm{SF}} = L^2/D_0 N^2 \propto 1/\phi^2} $, where $ L $ is the length of the periodic one-dimensional box and $ \phi = \sigma N/L $~\cite{Marchesoni2006}. 

By using the FACF-based version of the NC algorithm, the long-time tails at low densities can be obtained with high resolution. Figure~\ref{fig:3} shows the results for different $ \phi $. The theoretical predictions for the long-time behavior $ t \gg \tau_{\textrm{SF}} $~\cite{Marchesoni2006}
\begin{equation}
    Z_{\textrm{eq}}(t) =-\frac{1}{4 \sqrt{\pi}} \frac{D_0}{\tau_{\textrm{SF}}}\left(\frac{t}{\tau_{\textrm{SF}}}\right)^{-3/2},
 \label{eq:VACFSFD}
\end{equation}
are indicated by the solid lines. The results of the simulations match the long-time prediction of Ref.~\cite{Marchesoni2006}. As established in previous works~\cite{Marchesoni2006, Taloni2008}, the VACF curves collapse when rescaled by the single-file time scale: plotting $ {-Z_{\textrm{eq}}(t)\,\sigma^2/D_0^2\,\phi^2} $ versus ${t/{\tau_{\textrm{SF}}}}$ yields universal behavior (see inset of Fig.\ref{fig:3}). 

To highlight once again the improvements that can be obtained by the NC algorithm, the VACF obtained with the Frenkel algorithm~\cite{Frenkel1987}, which also suppresses noise by leaving out correlations between the initial velocity and the Brownian fluctuations (see Appendix~\ref{app:HS}), is depicted in Fig.~\ref{fig:3} with crosses. The FACF-based method resolves the long-time tail more clearly. The good agreement between our simulation results and the well-established theoretical prediction~\eqref{eq:VACFSFD} further confirms that cross-correlations vanish in equilibrium and that the effective FACF suffices to capture the single-file dynamics accurately.

\section{Nonequilibrium} \label{sec:NEQ}

To assess whether the NC algorithm remains valid in Brownian nonequilibrium systems, we first recall that in equilibrium, the VACF is equal to the negative of the FACF (up to a constant prefactor $\mu^2/d$). This identity, which culminates in Eq.~\eqref{eq:CCAREZERO}, makes the NC algorithm broadly applicable to many conventional soft-matter systems under equilibrium conditions. However, this relationship [expressed in Eq.~\eqref{eq:VACFEQ}], does not always hold in Brownian nonequilibrium systems, where cross-correlations between the full displacement and the force can become non-negligible. In these cases, the general result Eq.~\eqref{eq:VACFGENERALES} remains exact under the assumed stationarity and applicable to obtain valid results.

To substantiate these claims, we analyze a set of deliberately constructed example systems that are out of equilibrium by design. In this context, our aim is not to evaluate the performance or efficiency of the NC algorithm under such conditions. Rather, our goal is to provide a proof of concept that highlights the necessity of explicitly accounting for the cross-correlation term when treating Brownian nonequilibrium systems.

The simplicity of our example systems allows for an analytic evaluation of the corresponding cross-correlation term. Since Eq.~\eqref{eq:VACFGENERALES} remains valid even in Brownian nonequilibrium systems (as it only requires stationarity), it provides a practical strategy for extending the NC algorithm to such conditions. In particular, if the cross-correlation term can be computed or reasonably approximated analytically, the VACF may be obtained by first extracting the FACF from simulations and then correcting it using the analytically determined cross-correlation contribution. In this way, uncertainties arising from the explicit dependence on noise can be mitigated, even in Brownian nonequilibrium systems where cross-correlations play a non-negligible role.

\subsection{Driven particles}

Let us introduce a system where the tagged particle exhibits a constant nonzero average velocity, $ {\langle \dot{\vec{R}} \rangle = const.} $. This persistent drift breaks detailed balance and clearly places the system out of thermal equilibrium. For analytical convenience, we define the fluctuating force as $ {\Delta \vec{F}(t) = \vec{F}(t) - \langle \dot{\vec{R}} \rangle / \mu} $, and assume that its mean vanishes, i.\@e.\@, $ {\langle \Delta \vec{F}(t) \rangle = 0} $. Furthermore, we require that the autocorrelation of the fluctuating force decays to zero in the long-time limit $ {\lim\limits_{t\rightarrow\infty}\langle \Delta \vec{F}(t) \cdot \Delta \vec{F}(0) \rangle = 0} $ and that its correlation with the noise term also vanishes asymptotically, $ {\lim\limits_{t\rightarrow\infty} \langle \Delta \vec{F}(t + \tau_0) \cdot \boldsymbol{\eta}(\tau_0) \rangle = 0} $. These correlation functions are further assumed to decay sufficiently fast so that their time integrals remain finite as $t \rightarrow \infty$. Such assumptions are physically reasonable, as force fluctuations typically decorrelate on microscopic time scales.

Under these conditions, we can evaluate the cross-correlation term $ {\langle \Delta \vec{R}(t) \cdot \delta \vec{R}(t)\rangle} $ with $ {\Delta \vec{R}(t) = \delta \vec{R}(t) + \Delta \vec{R}_0(t)} $ by substituting Eqs.~\eqref{eq:deltaR} and \eqref{eq:DeltaR} for the displacements $\delta \vec{R}(t)$ and $\Delta \vec{R}_0(t) $. Noting that
$ {\langle \vec{F}(t) \cdot \boldsymbol{\eta}(t') \rangle = \langle \Delta \vec{F}(t) \cdot \boldsymbol{\eta}(t') \rangle} $ and $ {\langle \vec{F}(t) \cdot \vec{F}(t') \rangle = \langle \Delta \vec{F}(t) \cdot \Delta \vec{F}(t') \rangle + \langle \dot{\vec{R}}\rangle^2/\mu^2} $, we obtain 
\begin{align}
    \left\langle \Delta \vec{R}(t) \cdot \delta \vec{R}(t) \right\rangle  =& 2\mu^2 \int_0^t \int_0^s \left\langle \Delta \vec{F}(s') \cdot \Delta \vec{F}(0) \right\rangle \, \textrm{d} s' \, \textrm{d} s \nonumber \\ & + \mu \int_0^t \int_0^t \left\langle \Delta \vec{F}(s) \cdot \boldsymbol{{\eta}}(s') \right\rangle \, \textrm{d}s \, \textrm{d} s' \nonumber \\ & + \langle \dot{\vec{R}} \rangle^2 t^2. \label{eq:NECC}
\end{align}
Here, we use the fact that the autocorrelation of $\Delta \vec{F}(t)$ retains the same stationarity properties as $\vec{F}(t)$.

In the long-time regime, the two remaining integrands vanish while decaying sufficiently fast so that the two integrals contribute at most terms of order $O(t)$ by assumption. Therefore, the final term, which scales as $t^2$, dominates for long times, and we conclude that $ {\left\langle \Delta \vec{R}(t) \cdot \delta \vec{R}(t) \right\rangle = O(t^2)} $. Consequently, its time derivatives contribute at least at a constant of magnitude $ \propto \langle \dot{\vec{R}} \rangle $ to the VACF in the long-time regime, and neglecting the cross-correlation can lead to inaccurate or unreasonable approximations.

\begin{figure}[ht!]
\includegraphics[width=0.49\textwidth]{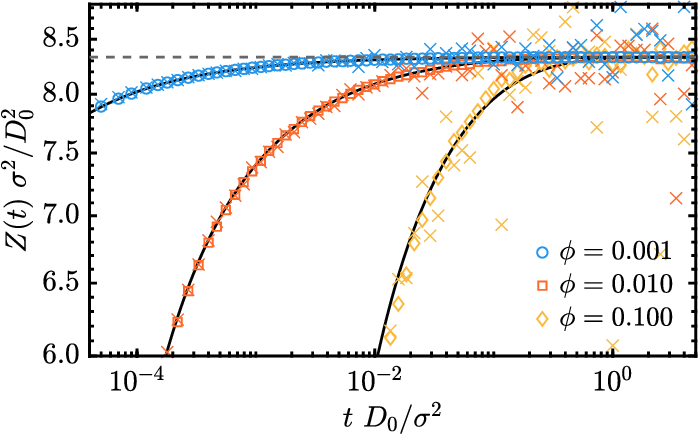}
\caption{Velocity autocorrelation function (VACF) for driven Brownian hard spheres at different volume fractions $ \phi $. The driving force is $ 
{\vec{F}_0 = (5,0,0)\, k_B T / \sigma} $ and the lines are analytic predictions. VACF results are calculated by evaluating the effective force autocorrelation function of the particle interactions and correcting the results with the analytic result for the cross-correlations (symbols) and the standard Frenkel algorithm~\cite{Frenkel1987} (crosses). The dashed line corresponds to the asymptotic value after correcting the NC expression with the finite cross-correlation term. See Appendix~\ref{app:HS} for the corresponding simulation parameters.}
\label{fig:4}
\end{figure}

As an illustrative example for such a system, consider a system in flow equilibrium with interacting particles that are all driven by the same external force, $\vec{F}_0 = \langle \dot{\vec{R}} \rangle / \mu$. In this scenario, the system is expected to behave like an equilibrium ensemble when viewed in a \qq{co-moving reference frame}, where $ \Delta \vec{F}(t) $, i.\@e.\ the interaction contribution, takes on the role of the total force $ \vec{F}(t) $ in the \qq{co-moving Langevin equation} for the corresponding velocity $ \dot{\vec{R}}(t) - \langle \dot{\vec{R}}\rangle $. Consequently, the first two terms of Eq.~\eqref{eq:NECC}, which correspond to the second time derivative of the cross-correlation term in the co-moving frame of reference, must cancel each other out [see Eq.~\eqref{eq:NFCF}]. If these contributions vanish, the general result for the VACF~\eqref{eq:VACFGENERALES} becomes
\begin{equation} 
Z(t) = -\frac{\mu^2}{d}\left\langle \Delta \vec{F}(t) \cdot \Delta \vec{F}(0) \right\rangle + \frac{\mu^2}{d} \vec{F}_0^2 ,
\label{eq:VACFDRIVEN} 
\end{equation} 
where we have again used the decomposition ${\langle \vec{F}(t) \cdot \vec{F}(t') \rangle = \langle \Delta \vec{F}(t) \cdot \Delta \vec{F}(t') \rangle + \langle \dot{\vec{R}} \rangle^2 / \mu^2}$. Since the autocorrelation of the force fluctuations decays to zero as $t \rightarrow \infty$, the cross-correlation term is, therefore, a dominant (constant) contribution to the VACF in the long-time limit and cannot be neglected.

To numerically validate that the cross-correlations are important for driven particles in flow equilibrium, we perform three-dimensional BD simulations of hard spheres at different volume fractions, which are driven by the force $ {\vec{F}_0 = (5,0,0) \, k_B T/ \sigma} $ (see Appendix~\ref{app:HS} for details and the simulation parameters). We calculate the VACF using an adjusted version of the NC algorithm based on the exact relation Eq.~\eqref{eq:VACFDRIVEN}: We calculate the correlation function $ {\langle \Delta \vec{F}(t) \cdot \Delta \vec{F}(0) \rangle} $ using the effective interaction forces and correct the results afterwards by adding $ \mu^2 \vec{F}_0^2 / d $ (symbols). For comparison, the crosses show the VACF obtained by the noise-suppressing Frenkel algorithm~\cite{Frenkel1987} from the same simulations, confirming the result from the adjusted NC expression (while being noisier). The lines correspond to analytic results for the dilute regime based on Eq.~\eqref{eq:VACFDRIVEN} and calculated by substituting Eq.~\eqref{eq:ZEQ} for $  -(\mu^2/d)\langle \Delta \vec{F}(t) \cdot \Delta \vec{F}(0) \rangle $. The VACF curves presented in Fig.~\ref{fig:4} converge to a constant (dashed line) that is only correct when incorporating the cross-correlation term $ \mu^2 \vec{F}_0^2 / d $.

\subsection{Active Brownian particle}

Another Brownian system which is inherently out of equilibrium is the active Brownian particle (ABP) in an external potential $ U(\vec{R}) $. In two dimensions, such a particle can be described by the Langevin equation~\eqref{eq:LE}, inserting 
\begin{equation}
\vec{F}(t) = F_{\textrm{eff}} \, \vec{u}(t) - \nabla_{\vec{R}} U(\vec{R}(t)),
\end{equation}
where $ F_{\textrm{eff}} $ denotes the magnitude of the effective propulsion force and $\vec{u}(t) = (\cos[\varphi(t)], \sin[\varphi(t)])^{\textrm{T}} $ specifies the ABP's orientation. The angle $\varphi(t)$ is a stochastic variable governed by the Langevin equation~\cite{Bechinger2016}
\begin{equation}
    \dot{\varphi}(t) = \eta_r(t),
\end{equation}
with $\eta_r(t) $ being a Gaussian white noise characterized by ${\langle \eta_r(t) \rangle = 0} $ and $ {\langle \eta_r(t)\eta_r(t') \rangle = 2D_r \delta(t - t')} $, where $D_r$ is the rotational diffusion coefficient. The rotational dynamics and the translational noise $\boldsymbol{\eta}(t)$ are assumed to be statistically independent. It can be readily verified that $ \vec{F}(t) $ satisfies the conditions outlined in Sec.~\ref{sec:Gen}.

Even in the absence of an external potential, i.\@e.\@, for $U(\vec{R}) \equiv 0$, the cross-correlation term of the VACF generally does not vanish, as can easily be checked: The well-established relation~\cite{Tao2005, Loewen2020} for the VACF of a free ABP is
\begin{align}
    Z(t) & = \frac{\mu^2}{2} F^2_{\textrm{eff}} \exp(-D_r t) = \frac{\mu^2}{2} \left\langle \vec{F}(t) \cdot \vec{F}(0) \right\rangle, 
 \label{eq:VACFABP}
\end{align}
for $t>0$. This expression differs by a sign from the result that would be expected for negligible cross-correlations. Consequently, the cross-correlation term must balance the discrepancy between the VACF and the FACF. 

Inserting the VACF of an ABP and the FACF given by $ {\langle \vec{F}(t) \cdot \vec{F}(0) \rangle = F_{\textrm{eff}}^2 \langle \vec{u}(t) \cdot \vec{u}(0) \rangle} $ in the general relation Eq.~\eqref{eq:VACFGENERALES}, we find
\begin{align}
    \frac{\textrm{d}^2}{\textrm{d}t^2}\left\langle \Delta \vec{R}(t) \cdot \delta \vec{R}(t) \right\rangle
    &= d\,Z(t) + \mu^2 \left\langle \vec{F}(t) \cdot \vec{F}(0) \right\rangle. \nonumber \\
    & = 2\mu^2 F^2_{\textrm{eff}} \exp(-D_r t)
\end{align}
for $ t > 0 $. This expression has the same lifetime $ 1/D_r $ as the FACF, which highlights that the cross-correlation term cannot be neglected even for arguably the simplest active system.

Next, we consider an ABP confined in a harmonic trap~\cite{Caraglio2022} --- a nontrivial example in which the original NC algorithm can be adjusted to yield highly accurate results from simulations. In this setting, the force acting on the particle is given by
\begin{equation}
    \vec{F}(t) = F_{\textrm{eff}} \, \vec{u}(t) - k \vec{R}(t),
\end{equation}
where the harmonic potential is defined as  $ {U(\vec{R}) = k \vec{R}^2 / 2}$. 

By using our general assumptions in Sec.~\ref{sec:Gen} regarding the FACF and by exploiting that the translational noise and $ \vec{u}(t) $ are uncorrelated, the cross-correlation term of the general VACF relation~\eqref{eq:VACFGENERALES} yields
\begin{align}
      \frac{\textrm{d}^2}{\textrm{d}t^2}  \langle \Delta \vec{R}(t) \cdot \delta \vec{R}(t) \rangle  &= 2 \mu^2 \left\langle \vec{F}(t) \cdot \vec{F}(0) \right\rangle  \nonumber \\ & -  \mu k \, \frac{\textrm{d}^2}{\textrm{d}t^2} \int_0^t \int_0^t \langle \vec{R}(s) \cdot \boldsymbol{\eta}(s') \rangle \, \textrm{d}s' \, \textrm{d} s.
\end{align}
The second term on the right-hand side can be calculated analytically using the general solution
\begin{equation}
    \vec{R}(t) = \vec{R}(0)\,\textrm{e}^{-\mu k t} + \int_0^t \textrm{e}^{-\mu k (t-s) }\, \left[\mu F_{\textrm{eff}} \, \vec{u}(s) + \boldsymbol{\eta}(s)\right] \, \textrm{d}s,
\end{equation}
and evaluating the noise averages. For $ t > 0 $, we obtain
\begin{equation}
     \frac{\textrm{d}^2}{\textrm{d}t^2} \int_0^t \int_0^t \langle \vec{R}(s) \cdot \boldsymbol{\eta}(s') \rangle \, \textrm{d}s' \, \textrm{d} s = 2 d D_0 \exp(-\mu k t),
\end{equation}
and, finally,
\begin{equation}
    Z(t) = \frac{\mu^2}{d} \left\langle \vec{F}(t) \cdot \vec{F}(0) \right\rangle - 2\mu k D_0\, \exp(-\mu k t),  \label{eq:ABFHP}
\end{equation}
for the corresponding VACF. This exact analytical result again demonstrates that approximating the VACF solely via the negative FACF, as is possible in the equilibrium, is insufficient in Brownian nonequilibrium systems. However, the second term in the VACF expression, which arises from the deterministic contribution of the harmonic trap, can be evaluated analytically. As it does not involve stochastic quantities, it is inherently noise-free. Hence, we can adjust the NC idea by calculating the FACF in simulations and then correcting the result afterwards by subtracting $  2\mu k D_0\, \exp(-\mu k t) $ to obtain maximum resolution.

\begin{figure}[ht!]
\includegraphics[width=0.49\textwidth]{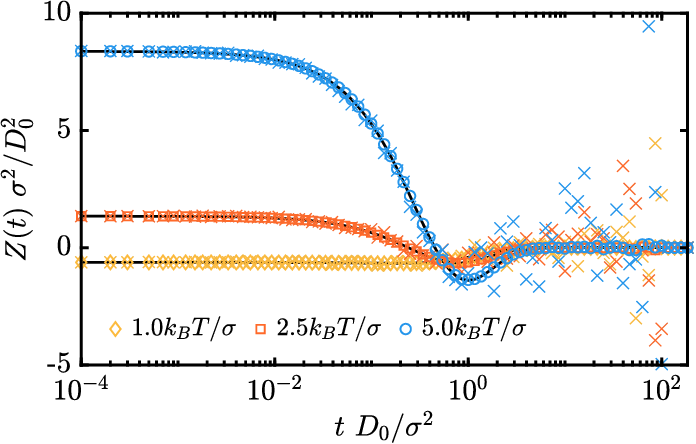}
\caption{Velocity autocorrelation function (VACF) of an active Brownian particle (ABP) in a two-dimensional harmonic potential with spring constant $ k = 1 k_B T / \sigma^2$ for different effective self-propulsion forces $ F_{\textrm{eff}} $. Symbols are calculated with the adjusted NC algorithm by evaluating the FACF during the simulations and analytically correcting the results afterwards. Crosses correspond to data obtained with a standard (noise suppressing) algorithm~\cite{Frenkel1987}. Lines are the theoretical predictions of Ref.~\cite{Caraglio2022}. See Appendix~\ref{app:ABPSIM} for the corresponding simulation parameters.}
\label{fig:5}
\end{figure}

Figure~\ref{fig:5} shows the VACF of an example system with $ {k = 1 k_B T / \sigma^2} $ for different propulsion strengths $ F_{\textrm{eff}} $. The symbols are obtained using Eq.~\eqref{eq:ABFHP} and the FACF obtained by BD simulations (see Appendix~\ref{app:ABPSIM} for details and the simulation parameters). The crosses are obtained using the noise-suppressing Frenkel algorithm for the VACF using the same simulation data. The results calculated via the adjusted NC algorithm resolve the VACF well, even in the regions where it has decayed to approximately zero. The results perfectly match theoretical predictions~\cite{Caraglio2022} (black lines in Fig.~\ref{fig:5}).

\section{Discussion} \label{sec:END}

In this work, we established the noise-cancellation (NC) algorithm as an exact and noise-suppressing method for computing the velocity autocorrelation function (VACF) in Brownian equilibrium systems that can be described by an overdamped Langevin equation of the form of Eq.~\eqref{eq:LE}. By deriving a general relation between the VACF, the force autocorrelation function (FACF), and displacement-force cross-correlations from the overdamped Langevin equation, we clarified the theoretical basis of the NC approach. We showed that in equilibrium, cross-correlation contributions vanish, allowing the VACF to be accurately obtained from the FACF alone, which can be efficiently computed during simulations. This reformulation not only unifies the NC algorithm with established methods to obtain long-time diffusion coefficients~\cite{Akesson1985} but also reveals its connection to the noise-suppressing Frenkel scheme~\cite{Frenkel1987} (see Appendix~\ref{app:Frenkel} for a detailed discussion), while demonstrating its superiority in terms of signal-to-noise ratio. For Brownian nonequilibrium systems that can be described by Eq.~\eqref{eq:LE}, where the cross-correlations do not vanish in general, we introduced an adjusted NC scheme that restores accuracy by analytically correcting for these contributions, extending the method's applicability to driven and active systems. Our simulations across soft and hard spheres in one, two, and three dimensions validate the effectiveness of both original and corrected NC approaches, resolving long-time behavior with distinct clarity.

Strictly speaking, our proof that the cross-correlation term of the VACF vanishes in equilibrium, which renders the NC algorithm exact, applies only to Brownian systems with continuous (\qq{soft}) interactions. However, we strongly anticipate that our result also holds for Brownian hard-sphere systems: all derivations remain valid for arbitrarily steep potentials, our example hard-sphere systems studied with the \qq{potential-free} BD algorithm suggest that the cross-correlations do not contribute, and previous EDBD simulations~\cite{Mandal2019} have empirically shown that cross-correlation terms are negligibly small. Furthermore, identifying the second time derivative of $\langle \delta \vec{R}^2(t) \rangle$ with the formal FACF expression defined in Ref.~\cite{Hanna1981} implies, per Hanna~\emph{et al.}~\cite{Hanna1981}, that the cross-correlations must vanish for hard spheres as well. Nonetheless, using an effective FACF to calculate the hard-sphere VACF (as done in this work) inevitably introduces an approximation whose accuracy depends on the quality of the effective force. This underscores the advantage of the original NC algorithm formulation~\cite{Mandal2019} for Brownian hard spheres, which computes the VACF directly from the derivatives of MSD without further approximations. Combined with our result that cross-correlation term vanishes, this makes the original NC algorithm a fully exact method for obtaining high-resolution VACF data from hard-sphere BD simulations in equilibrium.

Having demonstrated that the cross-correlation term of the VACF relation vanishes in thermal equilibrium but remains finite in nonequilibrium, we propose that it can act as a sensitive and practical indicator of nonequilibrium behavior in Brownian systems [with governing equation~\eqref{eq:LE}]. This term captures the correlation between the actual displacement of the system $\Delta \vec{R}(t)$ and the displacement $\delta \vec{R}(t)$ predicted from the net forces acting on it. In equilibrium, where the dynamics of Brownian systems are primarily governed by thermal fluctuations, this cross-correlation term vanishes, and the system's behavior is fully characterized by the FACF. In contrast, Brownian nonequilibrium systems are driven by persistent external forces, resulting in a significant correlation between the particle displacements and these forces. In such cases, the FACF alone is not enough to describe the system's dynamics. This provides a practical and quantitative criterion to distinguish equilibrium from nonequilibrium states and offers a robust framework for probing active and complex Brownian soft-matter systems.

Notably, our results can also be readily extended to colloidal equilibrium systems, where hydrodynamic interactions play a role. By modifying our derivations to include the position-dependent mobility matrices, we find that the cross-correlation term in the corresponding VACF should still vanish for such systems (see Appendix~\ref{app:HI}). This extension enables the NC algorithm to operate reliably across an even broader spectrum of soft‑matter systems, significantly expanding its practical reach.

The generality and efficiency of the NC algorithm position it as a powerful tool for a broad range of physical systems that can be modeled by overdamped Langevin equations. Its application can advance studies of active matter~\cite{Granek2022}, crowded and glassy environments~\cite{Ni2013,Mandal2020}, and systems with complex interactions, where precise long-time dynamics are critical yet hard to resolve. Moreover, the ability to derive various dynamical observables from the FACF and cross-correlation terms opens the door for data-efficient modeling~\cite{Asghar2024} in theoretical and experimental contexts. We anticipate that the NC algorithm will play a central role in future investigations of time-correlation functions, transport phenomena, and nonequilibrium dynamics in soft and active matter.

\begin{acknowledgments}
The authors thank Timo Knippenberg for helpful feedback and Regina Rusch for fruitful discussions. The computational results presented here have been achieved (in part) using the LEO HPC infrastructure of the University of Innsbruck. This research was funded in part by the Austrian Science Fund (FWF) 10.55776/P35673. We acknowledge the use of AI tools to polish, condense, and enhance our writing. 
\end{acknowledgments}

\appendix
\onecolumngrid

\section{Derivation of the VACF and FACF relation for equilibrium} \label{app:eq}

To show that Eq.~\eqref{eq:VACFEQ} is an exact identity in equilibrium, we begin by calculating the time derivative of the many-body MSD Eq.~\eqref{eq:MSDEQ}. Inserting the Smoluchowski equation~\eqref{eq:Smol} yields
\begin{align}
    \frac{\partial}{\partial t} \left\langle \Delta \vec{R}^2(t-t') \right\rangle_{\textrm{eq}} & = \int \!\int \textrm{d}\vec{r} \, \textrm{d}\vec{r}'\, \left( \vec{r}_1 - \vec{r}_1' \right)^2 \,
\left[ \hat{\Omega}_{\vec{r}} \mathds{P}( \vec{r},t\mid\vec{r}', t') \right] P_{\textrm{eq}}( \vec{r}') \nonumber \\
    & = \int \!\int \textrm{d}\vec{r} \, \textrm{d} \vec{r}' \, \left[  \hat{\Omega}_{\vec{r}}^{\dagger} \, \left( \vec{r}_1 - \vec{r}_1' \right)^2 \right] \,
    \mathds{P}( \vec{r},t\mid\vec{r}', t') P_{\textrm{eq}}( \vec{r}') \nonumber \\
    & = 2\mu \int \!\int \textrm{d}\vec{r} \, \textrm{d} \vec{r}' \,  \, \vec{F}_1(\vec{r}) \cdot \left( \vec{r}_1 - \vec{r}_1' \right) \,
    \mathds{P}(\vec{r},t\mid\vec{r}', t') P_{\textrm{eq}}(\vec{r}') + 2 D_0 d,
\end{align}
where we used integration by parts to replace the Smoluchowski operator with its adjoint [which is the operator~\eqref{eq:BACKOP} but acting on configuration $ \vec{r}=(\vec{r}_1,\ldots, \vec{r}_N) $]. Next, we calculate the derivative with respect to $t'$ of this relation and insert the Kolmogorov backward equation~\eqref{eq:KBE}. We find
\begin{align}
    \frac{\partial}{\partial t'} \frac{\partial}{\partial t} \left\langle \Delta \vec{R}^2(t-t') \right\rangle_{\textrm{eq}} & = -  2\mu \int \!\int\textrm{d}\vec{r} \, \textrm{d} \vec{r}' \,  \, \vec{F}_1(\vec{r}) \cdot \left( \vec{r}_1 - \vec{r}_1' \right) \,
    P_{\textrm{eq}}( \vec{r}')\left[ \hat{\Omega}^{\dagger}_{\vec{r}'} \mathds{P}(\vec{r},t\mid\vec{r}', t')\right]  \nonumber \\
    & = -  2\mu  \int\!\int \textrm{d}\vec{r} \, \textrm{d} \vec{r}' \,  \, \vec{F}_1(\vec{r}) \cdot \left( \vec{r}_1 - \vec{r}_1' \right) \,
     \hat{\Omega}_{\vec{r}'} \left[ \mathds{P}( \vec{r},t\mid\vec{r}', t')P_{\textrm{eq}}( \vec{r}') \right] \nonumber \\
    & = - 2\mu \int\!\int \textrm{d}\vec{r} \, \textrm{d} \vec{r}' \,  \, \left[ \hat{\Omega}_{\vec{r}'}^{\dagger} \,  \vec{F}_1(\vec{r}) \cdot \left( \vec{r}_1 - \vec{r}_1' \right) \right] \,
    \mathds{P}(\vec{r},t\mid\vec{r}', t')P_{\textrm{eq}}(\vec{r}') \nonumber \\
    & =  2\mu^2 \left\langle \vec{F}_1(r) \cdot \vec{F}_1(r') \right\rangle_{\textrm{eq}}
\end{align}
with $ \vec{r} = \vec{r}(t) $ and $ \vec{r}' = \vec{r}(t') $, as well as $ t > t' $. Here, we applied the identity $ P_{\textrm{eq}}(\vec{r}) \hat{\Omega}_{\vec{r}}^{\dagger}(\ldots) = \hat{\Omega}_{\vec{r}} [ (\ldots ) \, P_{\textrm{eq}}( \vec{r}) ] $ holding in equilibrium~\cite{Jones2003} and used integration by parts to obtain the result. In equilibrium, we know by definition that
\begin{equation}
    Z_{\textrm{eq}}(t-t') = - \frac{1}{2d} \frac{\partial}{\partial t} \frac{\partial}{\partial t'} \left\langle \Delta \vec{R}^2(t-t') \right\rangle_{\textrm{eq}} = -\frac{\mu^2}{d} \left\langle \vec{F}_1(t) \cdot \vec{F}_1(t') \right\rangle_{\textrm{eq}},
\end{equation}
which leads to the equilibrium VACF given in Eq.~\eqref{eq:VACFEQ}. Here, we follow our convention established above, where $ \vec{F}(t) $ is an abbreviation for $ \vec{F}(\vec{r}(t)) $, and we use $ \vec{F}(t') $ for $ \vec{F}(\vec{r}(t')) $. An alternative derivation for this identity in Laplace space can be found in Ref.~\cite{Hanna1981}.

\section{WCA Brownian dynamics simulation details} \label{app:WCA}

We perform conventional BD simulations~\cite{Ermak75, Ermak1978} of soft WCA particles. In these simulations, the position of the Brownian particles is updated by
\begin{equation}
    \vec{r}_i(t + \Delta t) = \vec{r}_i(t) + \frac{D_0}{k_B T} \vec{F}_i(t)\,\Delta t + \sqrt{2D_0 \Delta t}\,\vec{W}(t),
\end{equation}
where $ \Delta t $ is the size of the simulation step, $ T $ is the temperature, and $ k_B $ is the Boltzmann constant. The random contribution to the motion is modeled via the vector $ \vec{W}(t) $, whose components are standard normally distributed. This integration scheme results from applying the stochastic Euler method to the overdamped Langevin equation~\eqref{eq:LE}. In a physical context, this update rule is also known as the Ermak–McCammon algorithm~\cite{Ermak75, Ermak1978}.

All simulations are performed in units of the particle diameter $ \sigma $, the diffusion coefficient $ D_0 $, the Brownian time $ \sigma^2/D_0 $, and the thermal energy $ k_B T $. At the start of the simulations, $ N = 1000 $ particles are randomly placed in a cubic simulation box with box length $ L $ and periodic boundary conditions. The box length is adjusted to match the desired density. Note that it was shown in Ref.~\cite{Mandal2019} that $ N = 1000 $ is sufficient to ensure that finite-size effects are negligible when analyzing the VACF. For each simulation of a given parameter set, $2.0 \times 10^9$ simulation steps of size $\Delta t = 2.0 \times 10^{-4}\,\tau_D$ are performed. Before data collection for the correlation functions begins, the system is equilibrated for $20\,\tau_D$. To calculate the correlation functions \qq{on the fly} (i.e., during the simulations), we use the algorithm of Ref.~\cite{Siems2018}. We additionally average our results over 10 independent simulations with different noise realizations and initial conditions.

To directly calculate the VACF for comparison with the NC algorithm (crosses in Fig.~\ref{fig:1}), we approximate it via the displacement correlation function using $ {\langle \dot{\vec{R}}(t) \cdot \dot{\vec{R}}(0) \rangle \approx \langle [\vec{R}(t+\Delta t) - \vec{R}(t)] \cdot [\vec{R}(\Delta t) - \vec{R}(0)] \rangle / \Delta t^2} $. Technically, $ \dot{\vec{R}}(t) $ is not a well-defined observable for Brownian particles because we assume overdamped dynamics. Thus, $ \dot{\vec{R}}(t) $ should be understood as a formal symbol, and we always interpret it as the displacement per simulation step in the context of our simulations. For maximum statistics, the FACF and VACF are calculated by averaging the results over all particles of the system. The source code of the performed simulations is available at Ref.~\cite{Git}.

\section{Hard-sphere Brownian dynamics simulation details} \label{app:HS}

To perform BD simulations of hard spheres, we apply the \qq{potential-free} algorithm introduced by Heyes and Melrose~\cite{Heyes1993}, which assumes that only particle pairs interact at once (which is a good approximation for dilute systems). In this version of the BD algorithm, the particle positions are updated via~\cite{Foss2000}
\begin{equation}
    \vec{r}_i(t + \Delta t) = \vec{r}_i(t) + \Delta \vec{r}_i^{\textrm{HS}} + \sqrt{2D_0\Delta t} \vec{W}(t),
\end{equation}
where the last term is again the Brownian contribution and $ \Delta \vec{r}_i^{\textrm{HS}} $ corrects for the forbidden particle overlaps. In detail, during each simulation step, all particles are first displaced with respect to the Brownian contribution. Then, it is checked whether some spheres are overlapping. If two particles $ i $ and $ j $ overlap, their position is updated by $ \Delta \vec{r}_i^{\textrm{HS}} $ and $ \Delta \vec{r}_j^{\textrm{HS}} $ respectively, where these displacements shift the particle centers along their connection vector until the particles are in touch. The formulas for the displacements (that are only applied if $ r_{ij} < \sigma $) are given by~\cite{Foss2000}
\begin{align}
    \Delta \vec{r}_i^{\textrm{HS}} & = \frac{1}{2} (r_{ij} - \sigma) \frac{\vec{r}_{ij}}{r_{ij}} \\
    \Delta \vec{r}_j^{\textrm{HS}} & = -\frac{1}{2} (r_{ij} - \sigma) \frac{\vec{r}_{ij}}{r_{ij}}
\end{align}
with $ \vec{r}_{ij} = \vec{r}_j(t) - \vec{r}_i(t) $ and the corresponding absolute value $ r_{ij} = |\vec{r}_{ij}| $. The corresponding effective interaction forces $ \vec{F}_i(t) $ and $ \vec{F}_j(t) $ are approximated as discussed in Refs.~\cite{Sunol2023, Marenne2017, Foss2000} using Eq.~\eqref{eq:HSForce}.

Since the potential-free algorithm assumes that interactions occur exclusively between particle pairs, the sequence in which particles are tested for overlap is, in principle, inconsequential. However, in practice, the displacement of particles by $\Delta \vec{r}_i^{\textrm{HS}}$ can inadvertently introduce new collisions and unphysical overlaps with neighboring particles. To mitigate these effects, implementations often employ multiple iterations of overlap resolution within each time step. Nevertheless, the potential-free algorithm may introduce artifacts when attempting to resolve multi-particle collisions. In the present study, we focus on very dilute systems, where such events are expected to be rare. Accordingly, we perform a single overlap check per time step and argue that this is sufficient for an efficient proof of concept. For more detailed investigations, especially in denser regimes, multiple iterations of overlap resolution are recommended to ensure physical behavior.

All simulations are conducted using reduced units based on the particle diameter $\sigma$, the diffusion coefficient $D_0$, the thermal energy $k_B T$, and the Brownian time $\tau_D = \sigma^2 / D_0$. Across all studied systems, regardless of the dimensionality, we consistently use $N = 1000$ particles. For the one-dimensional setup, we apply a time step of $ 2.0 \times 10^{-2}\, \tau_D$,  $ 2.0 \times 10^{-2}\, \tau_D$, and  $ 2.6 \times 10^{-2}\, \tau_D$ for $ \phi = 5\% $, $ \phi=10\% $, and $ \phi = 20\% $, respectively. For the equilibrium simulations in three dimensions, the length of the time step $ \Delta t $ is set to $ 2.0 \times 10^{-4} \, \tau_D $ and $ 6.0 \times 10^{-4} \, \tau_D $ for $ \phi = 1\% $ and $ \phi = 10\% $. To perform the three-dimensional simulations involving driven particles, a smaller time step of $\Delta t = 1.0 \times 10^{-5}\, \tau_D$ is employed instead. At the beginning of each simulation, particles are randomly distributed within a simulation box with equal side lengths and periodic boundary conditions. Here, the side length of the box is adjusted to match the target density. Prior to computing correlation functions, all systems are equilibrated for $10^5$ time steps.

For the one-dimensional simulations at a density of $\phi = 5\%$, we perform $2.5 \times 10^9$ time steps and average the results over $10$ independent runs. At $\phi = 10\% $, a number of $ 2.0 \times 10^9 $ steps is used, and the averaging is performed over $ 5 $ simulations. For the higher tested density of $\phi = 20\%$, we conduct $ 5 $ simulations as well, each consisting of $1.0 \times 10^9$ steps.

In the three-dimensional equilibrium simulations, results are averaged over $10$ independent runs. For these setups, we employ $2.5 \times 10^9$ time steps at a volume fraction of $\phi = 1\%$, and $1.3 \times 10^9$ steps for $\phi = 10\%$. In simulations involving driven particles, an additional displacement $ \Delta \vec{R}_F = D_0 \vec{F}_0 \Delta t / k_B T $ is applied at each time step to account for the external driving force, coinciding with the update of the positions due to the Brownian motion. For these driven cases, we perform a single simulation per volume fraction, each consisting of $5.0 \times 10^8$ steps to obtain the presented data.

To calculate the correlation functions during the simulations, the algorithm described in Ref.~\cite{Siems2018} is applied. With this, an approximation for the hard-sphere FACF can be calculated using the approximated interaction forces $ \vec{F}_i(t) $. Note that the results match analytical results for hard spheres as well as approximations using the WCA potential, which shows that the approach of Refs.~\cite{Heyes1993, Foss2000} is sufficient to calculate FACF data from hard-sphere BD simulations (at least in the resolution of the proof of concept presented in this work). 

To calculate the VACF with an alternative algorithm for reasons of comparison, we apply the noise-suppressing Frenkel algorithm~\cite{Frenkel1987}. It argues that the Brownian contributions at time $ t $ are not correlated with the initial velocity and neglects the corresponding term in the VACF calculation~\cite{Frenkel1987}. The VACF is then computed by Eq.~\eqref{eq:VACFNOISE}, where the terms on the right-hand side are calculated numerically. Note that this method is exact in both equilibrium and nonequilibrium. Also, it has a significantly better resolution than calculating the VACF from the particle displacement correlations (as done for the soft spheres) since it eliminates one explicit dependence on the Brownian noise. More information on this method is given in Appendix~\ref{app:Frenkel}. The source code of the performed simulations is available at \cite{Git}.

\section{Active Brownian particle simulations} \label{app:ABPSIM}

To obtain the FACF from BD simulations of ABPs in a harmonic potential, we numerically integrate the corresponding equations of motion with the stochastic Euler method and obtain the updated rules
\begin{align}
    \vec{R}(t+\Delta t) & = \left[1-\frac{D_0}{k_B T}k \Delta t\right] \, \vec{R}(t) + \frac{D_0}{k_B T} F_{\textrm{eff}} \, \left(\cos[\varphi(t)], \sin[\varphi(t)]\right)^{\textrm{T}} \Delta t + \sqrt{2D_0 \Delta t} \vec{W}(t) \\
    \varphi(t+ \Delta t) & = \varphi(t) + \sqrt{2D_r\Delta t} W_r(t).
\end{align}
Here, $ W_r(t) $ is a standard normally distributed random number, and $ D_r = 3 D_0 / \sigma^2 $ is the rotational diffusion coefficient of a sphere. The initial position is set to $ \vec{R}(0) = 0 $ and the start orientation is chosen randomly. The corresponding FACF is calculated using the algorithm of Ref.~\cite{Siems2018} and $1000$ independent simulations of $ 2.0\times 10^6 $ steps of size $ \Delta t = 1.0 \times10^{-4} \, \tau_D $ for each analyzed effective propulsion force (where the system is always \qq{equilibrated} for $ 10^6 $ of the $ 2.0\times 10^6 $ steps before the calculation of the correlation functions starts). The direct data for the VACF is calculated in the same way as for the hard-sphere simulations~\cite{Frenkel1987}. The source code of the performed simulations is available at \cite{Git}.

\section{Connection to Frenkel algorithm} \label{app:Frenkel}

The foundation of the NC algorithm of splitting the particle displacements in the Brownian motion and the contributions of the collisions~\cite{Mandal2019} is based on the Frenkel algorithm~\cite{Frenkel1987}. In this preceding computation scheme (that was originally formulated on a lattice), the velocity $ \dot{\vec{R}}(t) $ of the tracer particle (and not the displacement) is divided into a free particle contribution $ \dot{\vec{R}}^{\textrm{free}}(t) $ and a part that contains all the information on the collisions and the acting forces $ \dot{\vec{R}}^{\textrm{int}}(t) $ (where we ignore the suggested bit-vector operation optimizations tailored to the hopping dynamics model used in Ref.~\cite{Frenkel1987} while calculating the cross-correlations in this work). Because the free particle velocity $ \dot{\vec{R}}^{\textrm{free}}(t) $ does not have a memory of the initial value $ \dot{\vec{R}}(0) $ for $ t > 0 $~\cite{Frenkel1987}, the VACF then simplifies to
\begin{equation}
    Z(t) = \frac{1}{d}\langle \dot{\vec{R}}(t) \cdot \dot{\vec{R}}(0) \rangle = \frac{1}{d}\langle \dot{\vec{R}}^{\textrm{int}}(t) \cdot \dot{\vec{R}}(0) \rangle, \label{eq:DEFFRENKEL}
\end{equation}
which also reduces the explicit noise dependence of the correlation function and, thus, the necessary sample size to obtain meaningful data in numerical computations, while still being an exact method (in equilibrium and nonequilibrium). When calculating the VACF via the Frenkel algorithm, $ \langle \dot{\vec{R}}^{\textrm{int}}(t) \cdot \dot{\vec{R}}(0) \rangle $ is the quantity which is determined numerically.

Although Ref.~\cite{Mandal2019} suggests that the NC scheme improves upon the traditional Frenkel algorithm, a direct comparison has not been possible yet. By reformulating the NC scheme as presented in this work, we can now directly contrast the two approaches based on their definitions. Identifying $ \dot{\vec{R}}^{\textrm{int}}(t) = \mu \vec{F}(t) $, the Frenkel algorithm is equal to the correlation function 
\begin{equation}
   \frac{\mu}{d} \left\langle \vec{F}(t) \cdot \dot{\vec{R}}(0) \right\rangle = -\frac{\mu^2}{d} \left\langle \vec{F}(t) \cdot \vec{F}(0) \right\rangle + \frac{1}{d} \frac{\textrm{d}^2}{\textrm{d}t^2}  \left\langle \Delta \vec{R}(t) \cdot \delta \vec{R}(t) \right\rangle. \label{eq:FRENKEL}
\end{equation}
which can be derived from Eq.~\eqref{eq:FNCF} after inserting $ \dot{\vec{R}}(0) = \mu\vec{F}(0) + \boldsymbol{\eta}(0) $ on the left-hand side. In equilibrium, where cross-correlations vanish, both the Frenkel and NC algorithms are, therefore, formally exact and, in principle, yield identical results. However, because
\begin{align}
    \langle \dot{\vec{R}}^{\textrm{int}}(t) \cdot \dot{\vec{R}}(0) \rangle & = \mu^2 \left\langle \vec{F}(t) \cdot \vec{F}(0) \right\rangle  + \mu \left\langle \vec{F}(t+\tau_0) \cdot \boldsymbol{\eta}(\tau_0)\right\rangle,
\end{align}
the Frenkel algorithm contains an explicit dependence on the Brownian fluctuations, and it tends to produce noisier data in simulations (here, we disregard the bit-vector optimization techniques proposed in Ref.~\cite{Frenkel1987} for hopping dynamics). This behavior is confirmed numerically in Fig.~\ref{fig:3}, where the Frenkel algorithm is used to compute the VACF directly (crosses). In equilibrium systems, the NC algorithm is therefore preferable due to its enhanced accuracy.

In Brownian nonequilibrium systems, however, the original NC formulation breaks down, while the Frenkel algorithm remains valid and yields numerically exact results. This robustness can be advantageous when analytical corrections to the NC scheme are not feasible. If such corrections are available, as demonstrated for active Brownian particles in a harmonic potential, the NC algorithm can outperform the Frenkel method (for instance, see the crosses in Fig.~\ref{fig:5}).

\section{Hydrodynamic interactions} \label{app:HI}

We consider a suspension of $N$ particles with positions $\vec{r} = (\vec{r}_1, \dots, \vec{r}_N)$ in thermal equilibrium, where the particles interact through hydrodynamic interactions described by a set of symmetric mobility tensors $\underline{\underline{\mu_{ij}}}(\vec{r})$. Each particle is subject to a conservative force $\vec{F}(\vec{r}) = (\vec{F}_1(\vec{r}), \dots, \vec{F}_N(\vec{r}))$ derived from the potential $U(\vec{r})$. Because of the hydrodynamic interactions, a force acting on any one particle affects the motion of all the others.

Based on this setup, the equation of motion for particle $i$ can be written (in It\^{o} form to be consistent with the stochastic Euler integration) as
\begin{equation}
    \dot{\vec{r}}_i(t) =  k_B T\sum_{j=1}^N \nabla_{\vec{r}_j} \cdot \underline{\underline{\mu_{ij}}}(\vec{r}(t)) + \sum_{j=1}^N \underline{\underline{\mu_{ij}}}(\vec{r}(t))\,\vec{F}_j(\vec{r}(t)) + \boldsymbol{\eta}_i(t),
\end{equation}
where we define the $ \alpha $-component of the matrix divergence via $ \left(\nabla_{\vec{r}_j} \cdot \underline{\underline{O}}\right)_\alpha = \sum_{\beta=1}^3 \frac{\partial O_{\alpha\beta}}{\partial r_{j\beta}} $ for a symmetric matrix $ \underline{\underline{O}} = (O_{\alpha \beta}) $ and the components $ r_{j\beta} $ of $ \vec{r}_j $. The noise satisfies the properties $\langle \boldsymbol{\eta}_i(t) \rangle_{\textrm{noise}} = 0$ and $\langle \boldsymbol{\eta}_i(t) \otimes \boldsymbol{\eta}_j(t') \rangle_{\textrm{noise}} = 2 k_B T\, \underline{\underline{\mu_{ij}}}(\vec{r}(t))\, \delta(t - t')$. Notably, this equation of motion contains a spurious drift term $ k_B T \sum_j \nabla_{\vec{r}_j} \cdot \underline{\underline{\mu_{ij}}}(\vec{r})$, since the noise is multiplicative due to the position dependence of the hydrodynamic mobility matrices. Without loss of generality, we choose $\vec{R}(t) = \vec{r}_1(t)$ as the position of the tracer particle. The diffusion tensors are given by $\underline{\underline{D_{ij}}}(\vec{r}) = k_B T\, \underline{\underline{\mu_{ij}}}(\vec{r})$, and the translational diffusion coefficient follows as $D_0 := \left\langle\mathrm{Tr}[\underline{\underline{D_{ii}}}(\vec{r})]\right\rangle_{P_\textrm{eq}}/d = k_B T\, \left\langle\mathrm{Tr}[\underline{\underline{\mu_{ii}}}(\vec{r})]\right\rangle_{P_\textrm{eq}}/d$, where $ \langle \cdot \rangle_{P_\textrm{eq}} $ is the average with respect to the stationary probability density $ P_{\textrm{eq}}(\vec{r}) $.

We summarize the deterministic contributions to the tagged particle’s equation of motion by introducing the generalized force $ \mathbfcal{{F}}(\vec{r}) = (\mathbfcal{{F}}_1(\vec{r}), \dots, \mathbfcal{{F}}_N(\vec{r}) )$, where $\mathbfcal{{F}}_{j}(\vec{r}) \equiv \mathbfcal{{F}}_{1j}(\vec{r}) $ and $ \mathbfcal{{F}}_{ij}(\vec{r}) := \left[k_B T \,\nabla_{\vec{r}_j} \cdot \underline{\underline{\mu_{ij}}}(\vec{r})
+ \underline{\underline{\mu_{ij}}}(\vec{r})\,\vec{F}_j(\vec{r})\right] / \mu$ and $ \mu = D_0 / k_B T $. We then define the deterministic and noise-induced displacements as $\delta \vec{R}(t)
= \mu \int_0^t \sum_j \mathbfcal{{F}}_j(s)\,\mathrm{d}s$
and $\Delta \vec{R}_0(t) = \int_0^t \boldsymbol{\eta}_1(s)\,\mathrm{d}s$, where we write $ \mathbfcal{{F}}_j(t) $ for $ \mathbfcal{{F}}_j(\vec{r}(t)) $ for convenience. With these definitions, the MSD decomposes as
\begin{equation}
    \langle \Delta \vec{R}^2(t) \rangle
    = \langle \Delta \vec{R}_0^2(t) \rangle
    - \langle \delta \vec{R}^2(t) \rangle
    + 2 \langle \Delta \vec{R}(t) \cdot \delta \vec{R}(t) \rangle,
\end{equation}
where in this appendix the full average is given by first taking the noise average $ \langle \cdot\rangle_{\textrm{noise}} $ and then averaging over the stationary probability density $ P_{\textrm{eq}}(\vec{r}) $. The first term evaluates to
\begin{equation}
    \langle \Delta \vec{R}_0^2(t) \rangle
    = \int_0^t \int_0^t
      \langle \boldsymbol{\eta}_1(s) \cdot \boldsymbol{\eta}_1(s') \rangle
      \,\mathrm{d}s\,\mathrm{d}s'
    = 2 k_B T \int_0^t \int_0^t
      \left\langle\mathrm{Tr}[\underline{\underline{\mu_{11}}}(\vec{r})]\right\rangle_{P_\textrm{eq}}\,\delta(s - s')
      \,\mathrm{d}s\,\mathrm{d}s'
    = 2 d D_0 t,
\end{equation}
which is identical to the noise-only result obtained in the absence of hydrodynamic interactions. Thus, the MSD continues to satisfy Eq.~\eqref{eq:MSD}.

Since we are considering an equilibrium system, the correlation functions $\langle \mathbfcal{{F}}_j(t) \cdot \mathbfcal{{F}}_k(t') \rangle$ must be stationary. Consequently, the mean-squared displacement arising from the generalized forces can be written as usual as
\begin{equation}
    \langle \delta \vec{R}^2(t) \rangle
    = \mu^2\int_0^t \int_0^t
      \sum_{j,k}
      \langle \mathbfcal{{F}}_j(s) \cdot \mathbfcal{{F}}_k(s') \rangle
      \,\mathrm{d}s\,\mathrm{d}s' =2 \mu^2\int_0^t \int_0^s
      \sum_{j,k}
      \langle \mathbfcal{{F}}_j(s') \cdot \mathbfcal{{F}}_k(0) \rangle
      \,\mathrm{d}s'\,\mathrm{d}s.
\end{equation}
From this expression, we obtain
\begin{equation}
    Z(t)
    = -\frac{\mu^2}{d}
      \sum_{j,k}
      \langle \mathbfcal{{F}}_j(t) \cdot \mathbfcal{{F}}_k(0)\rangle
      + \frac{1}{d}\,\frac{\mathrm{d}^2}{\mathrm{d}t^2}
        \langle \Delta \vec{R}(t) \cdot \delta \vec{R}(t) \rangle,
\end{equation}
for $ t > 0 $, by calculating the time derivatives of the MSD. This result is equivalent to Eq.~\eqref{eq:VACFGENERALES}.

We can also show that the cross-correlation term in this expression vanishes by adapting the derivation in Appendix~\ref{app:eq}. As a first step, we introduce the $N$-particle Smoluchowski operator for a system with hydrodynamic interactions, acting on a function $A = A(\vec{r})$ of the final configuration. It is defined as~\cite{Dhont1996}
\begin{equation}
    \hat{\Omega}_{\vec{r}} \, A
    = \sum_{i,j} \left[
        \left(\nabla_{\vec{r}_i} \otimes \nabla_{\vec{r}_j}\right) : \left(\underline{\underline{D_{ij}}}(\vec{r}) \, A\right)
        - \mu\,\nabla_{\vec{r}_i} \cdot \big( A\,\mathbfcal{{F}}_{ij} (\vec{r}) \big)
    \right],
\end{equation}
where the double contraction of two matrices $\underline{\underline{O}} = (O_{\alpha\beta})$ and $\underline{\underline{P}} = (P_{\alpha\beta})$ is defined by
$\underline{\underline{O}} : \underline{\underline{P}} = \sum_{\alpha\beta} O_{\alpha\beta} P_{\alpha\beta}$.
Likewise, the adjoint operator $\hat{\Omega}_{\vec{r}'}^{\dagger}$ acting on a function $B = B(\vec{r}')$ of the initial configuration is
\begin{equation}
    \hat{\Omega}_{\vec{r}'}^{\dagger} \, B
    = \sum_{i,j} \left\{
        \underline{\underline{D_{ij}}}(\vec{r}') : \left[\left(\nabla_{\vec{r}'_i} \otimes \nabla_{\vec{r}'_j} \right)\, B\right]
        + \mu\,\mathbfcal{{F}}_{ij}(\vec{r}') \cdot \nabla_{\vec{r}'_i} B
    \right\}.
\end{equation}
This operator serves as the generator of the dynamics for observables. Since the system is in equilibrium, the full average is identical to averaging over the propagator $  \mathds{P}(\vec{r}, t \mid \vec{r}', t') $, which solves the corresponding Smoluchowski and Kolmogorov backward equations, before taking the average with respect to $ P_{\textrm{eq}}(\vec{r}') $.

We can now evaluate the VACF from the MSD using the operators defined above. We begin with
\begin{align}
    \frac{\partial}{\partial t}
    \left\langle \Delta \vec{R}^2(t - t') \right\rangle
    &= \int\!\int \mathrm{d}\vec{r}\,\mathrm{d}\vec{r}'\,
      \left[\hat{\Omega}_{\vec{r}}^{\dagger}\, \left( \vec{r}_1 - \vec{r}_1' \right)^2\right]\,
        \mathds{P}(\vec{r}, t \mid \vec{r}', t')
       P_{\mathrm{eq}}(\vec{r}')
    \\
    &= 2\mu \sum_j  \int\!\int \mathrm{d}\vec{r}\,\mathrm{d}\vec{r}'\,
       \mathbfcal{{F}}_j(\vec{r}) \cdot
       \left( \vec{r}_1 - \vec{r}_1' \right)\,
       \mathds{P}(\vec{r}, t \mid \vec{r}', t')\,P_{\mathrm{eq}}(\vec{r}')
       + 2\,\left\langle \mathrm{Tr}[\underline{\underline{D_{11}}}(\vec{r})] \right\rangle_{P_\textrm{eq}}.
\end{align}
Note that the adjoint operator acts on $ \vec{r} $ in this step. Taking the derivative with respect to the initial time $t'$, we obtain
\begin{align}
    \frac{\partial}{\partial t'}\frac{\partial}{\partial t}
    \left\langle \Delta \vec{R}^2(t - t') \right\rangle
    &= -2\mu \sum_j\int\!\int \mathrm{d}\vec{r}\,\mathrm{d}\vec{r}'\,
        \left[
       \hat{\Omega}^{\dagger}_{\vec{r}'}\,
       \mathbfcal{{F}}_j(\vec{r}) \cdot
       \left( \vec{r}_1 - \vec{r}_1' \right)
       \right]\,
       \mathds{P}(\vec{r}, t \mid \vec{r}', t')\,P_{\mathrm{eq}}(\vec{r}')
    \nonumber \\
    &= 2\mu^2 \sum_{j,k}
       \left\langle
       \mathbfcal{{F}}_j(\vec{r}) \cdot
       \mathbfcal{{F}}_k(\vec{r}')
       \right\rangle
\end{align}
with $ \vec{r} = \vec{r}(t) $ and $ \vec{r}' = \vec{r}(t') $, as well as $ t > t' $. Here, we again used the identity $ P_{\mathrm{eq}}(\vec{r})\,\hat{\Omega}^{\dagger}_{\vec{r}}(\ldots) = \hat{\Omega}_{\vec{r}}\!\left[ (\ldots)\,P_{\mathrm{eq}}(\vec{r}) \right] $, which must remain valid because the equilibrium distribution is unchanged by the inclusion of hydrodynamic interactions and is still given by the Boltzmann distribution
$P_{\mathrm{eq}}(\vec{r}) \propto \exp[-U(\vec{r})/k_B T]$. Inserting the above expression into the definition of the VACF yields
\begin{equation}
    Z(t-t')
    = -\frac{1}{2d}
      \frac{\partial}{\partial t'}\frac{\partial}{\partial t}
      \left\langle \Delta \vec{R}^2(t - t') \right\rangle
    = -\frac{\mu^2}{d} \sum_{j,k}
      \left\langle
      \mathbfcal{{F}}_j(t) \cdot
      \mathbfcal{{F}}_k(t')
      \right\rangle.
\end{equation}
Thus, the cross-correlation term indeed vanishes in equilibrium and the VACF is fully determined by the autocorrelation function of the generalized forces. The NC algorithm therefore remains exact for equilibrium systems, even when hydrodynamic interactions are included.

Notably, for dilute systems where far-field hydrodynamics, described at the Oseen or Rotne-Prager-Yamakawa level, are valid, the VACF reduces to
\begin{equation}
    Z(t) = -\frac{1}{d} \sum_{j,k}
      \left\langle
      \underline{\underline{\mu_{1j}}}(t) \, \vec{F}_j(t) \cdot
      \underline{\underline{\mu_{1k}}}(0) \, \vec{F}_k(0)
      \right\rangle, \label{eq:VACFHI}
\end{equation}
which has the same structure as Eq.~\eqref{eq:VACFEQ}. This simplification arises because the Oseen and Rotne-Prager-Yamakawa mobility tensors are divergence-free (i.\@e.\@, $ \nabla_{\vec{r}_j} \cdot \underline{\underline{\mu_{ij}}}(\vec{r}) = 0 $)~\cite{Ermak1978}. Equation~\eqref{eq:VACFHI} is also consistent with the results of Fixman~\cite{Fixman1981, Liu2003}, who derived an equivalent expression for the self-diffusion coefficient.

\twocolumngrid

\end{document}